\documentclass{emulateapj}
\usepackage{nicefrac}
\usepackage{graphicx}
\usepackage{txfonts}
\usepackage{natbib}
\usepackage[scaled]{helvet}
\usepackage{epsfig}
\usepackage{url}
\usepackage[normalem]{ulem}
\usepackage{color}
\bibpunct{(}{)}{;}{a}{}{,}
\begin{document}

\title{\bf Photometric Variability of the B\MakeLowercase{e} Star Population}

\author{
Jonathan Labadie-Bartz$^{1}$,
Joshua Pepper$^{1}$, 
M. Virginia McSwain$^{1}$, 
J. E. Bjorkman$^{2}$, 
K. S. Bjorkman$^{2}$,
Michael B. Lund$^{3}$, 
Joseph E. Rodriguez$^{6,3}$, 
Keivan G. Stassun$^{3}$, 
Daniel J. Stevens$^{4}$,
B. Scott Gaudi$^{4,5}$,
David J. James$^{7,8}$, 
Rudolf B. Kuhn$^{9,10}$,
Robert J. Siverd$^{11}$,
Thomas G. Beatty$^{12,13}$
}

\affil{$^1$Department of Physics, Lehigh University, 16 Memorial Drive East, Bethlehem, PA 18015, USA}
\affil{$^2$Ritter Observatory, Department of Physics \& Astronomy, University of Toledo, 2801 W. Bancroft, Toledo, OH 43606-3390}
\affil{$^3$Department of Physics and Astronomy, Vanderbilt University, Nashville, TN 37235, USA}
\affil{$^4$Department of Astronomy, The Ohio State University, 140 W. 18th Ave., Columbus, OH 43210, USA}
\affil{$^5$Jet Propulsion Laboratory, California Institute of Technology, 4800 Oak Grove Dr, Pasadena, CA 91109}
\affil{$^6$Harvard-Smithsonian Center for Astrophysics, 60 Garden St, Cambridge, MA 02138, USA}
\affil{$^7$Cerro Tololo Inter-American Observatory, Casilla 603 La Serena, Chile}
\affil{$^8$Department of Astronomy, University of Washington, Box 351580, Seattle, WA 98195}
\affil{$^9$Southern African Large Telescope, P. O. Box 9, Observatory 7935, Cape Town, South Africa}
\affil{$^{10}$South African Astronomical Observatory, P. O. Box 9, Observatory 7935, Cape Town, South Africa}
\affil{$^{11}$Las Cumbres Observatory Global Telescope Network, 6740 Cortona Drive, Suite 102, Santa Barbara, CA 93117, USA}
\affil{$^{12}$Department of Astronomy \& Astrophysics, The Pennsylvania State University, 525 Davey Lab, University Park, PA 16802}
\affil{$^{13}$Center for Exoplanets and Habitable Worlds, The Pennsylvania State University, 525 Davey Lab, University Park, PA 16802}

\shorttitle{Photometric Variability of Be Stars}

\begin{abstract}
Be stars have generally been characterized by the emission lines in their spectra, and especially the time variability of those spectroscopic features. They are known to also exhibit photometric variability at multiple timescales, but have not been broadly compared and analyzed by that behavior. We have taken advantage of the advent of wide-field, long-baseline, and high-cadence photometric surveys that search for transiting exoplanets to perform a comprehensive analysis of brightness variations among a large number of known Be stars. The photometric data comes from the KELT transit survey, with a typical cadence of 30 minutes, baseline of up to ten years, photometric precision of about 1\%, and coverage of about 60\% of the sky. We analyze KELT light curves of 610 known Be stars in both the Northern and Southern hemispheres in an effort to study their variability. Consistent with other studies of Be star variability, we find most of the stars to be photometrically variable. We derive lower limits on the fraction of stars in our sample that exhibit features consistent with non-radial pulsations (25$\%$), outbursts (36$\%$), and long term trends in the circumstellar disk (37$\%$), and show how these are correlated with spectral sub-type. Other types of variability, such as those owing to binarity, are also explored. Simultaneous spectroscopy for some of these systems from the Be Star Spectral Database (BeSS) allow us to better understand the physical causes for the observed variability, especially in cases of outbursts and changes in the disk.

\end{abstract}

\section{Introduction}
Classical Be stars are a rapidly rotating subset of non-giant B-type stars. Unlike normal B-type stars, Be stars exhibit line emission (hence the `e' in Be), which is attributed to a gaseous circumstellar disk in Keplerian motion. The circumstellar disk of a Be star is best described by the viscous decretion disk model \citep[e.g.][]{Lee1991,Carciofi2011}, where the disk is formed and subsequently sustained by mass ejected from the stellar surface in discrete events called `outbursts' (e.g. \citet{Kroll1997} or \citet{Kee2014} for a theoretical framework, or \citet{Rivinius1998} or  \citet{Grundstrom2011} for observations). This ejected material then orbits the star, settling into a disk. However, if the flow of material from star to disk stops, the disk will dissipate through viscous forces, which facilitate the transfer of angular momentum to the outer disk (simultaneous with a loss of angular momentum in the inner disk), causing the disk to clear from the inside outward \citep{Haubois2012}, or alternatively via line-driven ablation \citep{Kee2016}, which also would clear the disk from the inside outward. The details of the physical mechanism which launches stellar material into orbit remains elusive, but very rapid (near critical) stellar rotation combined with non-radial pulsations (NRP) is theorized to play an important role in the mass transfer mechanism \citep{Rivinius2013}. The physical changes taking place within Be stars and their disks systems leave imprints in various observables that are accessible through spectroscopy, photometry, polarimetry, and interferometry, allowing the underlying mechanisms of variability to be studied through a variety of techniques.

Observations of Be stars during both both variable and non-variable phases provide insight into the physical mechanisms taking place, helping work towards a more thorough understanding of the `Be phenomenon'. Here we present long time-baseline, high-precision photometric observations of large numbers (hundreds) of Galactic Be stars, substantially increasing our knowledge of the photometric variability of this population.

Be stars are known to show variability in their brightness and spectral features across a large range of timescales from hours to decades. When variability in a Be star is observed, the associated timescales often give insight into the physical cause of these changes. Periodicity on shorter timescales of hours to days are typically attributed to stellar NRP. Outbursts and quasi-periodic oscillations are typically found on intermediate timescales of days to months, although outbursts occasionally have durations of years. The longest timescales typically involve changes in the disk, the most dramatic of which is the total disappearance (or reappearance) of the central star's circumstellar disk. It is not uncommon for these disks to appear and disappear over the course of years to decades \citep{McSwain2009}, and as such an object classified as a Be star must have shown emission at some point in time, but does not necessarily show emission (or posses a disk) in the current epoch. Other changes in the circumstellar environment and the disk can manifest themselves in a variety of observational ways. 

NRP are commonly observed in Be stars, with typical timescales between $\lesssim$0.1 day to 2 days. \citet{Cuypers1989} detect NRP in $\sim$82\% of a sample of 17 Be stars. In a sample of 57 Be stars, \citet{Gutierrez-Soto2007} detect short-term variability indicative of NRP in 74$\%$ of early-type Be stars, and in 31$\%$ of mid- and late-type Be stars. The photometric amplitudes associated with NRP in Be stars can be quite low, down to the sub-mmag level \citep{Emilio2010, Walker2005, Walker2005a, Saio2007}. This is further complicated by the fact that pulsational frequencies and amplitudes can shift, especially when the star also undergoes outbursts \citep{Hubert2007,Huat2009}. Although signatures of these pulsations can be very difficult to detect, all Be stars that have been analyzed with high-cadence, long-duration space-based photometry have been reported to be multiperiodic and to pulsate, with amplitudes decreasing with later spectral subtypes \citep{Rivinius2013}. It therefore seems that, as a class of objects, Be stars are pulsators.

Periodic variability has been detected in Be stars with periods longer than what can be explained by single NRP modes. Cyclical variability between 60 - 100 days was detected in the Be star $\delta$ Scorpii \citep{Jones2013}. \citet{Sterken1996} find periodic and quasi-periodic oscillations (QPO) in brightness in 4 Be stars with periods ranging between 4 and 93 days. In a similar analysis, \citet{Mennickent1994} detect QPO in two stars (27 CMa and 28 CMa) with periods between 10 and 20 days. \citet{Hubert1998} make use of Hipparcos photometry \citep{VanLeeuwen1997} and find QPO with a period of 11.546 days in the Be star MX Pup. One possible explanation or this type of periodic variability is the beating of two or more NRP modes with closely spaced frequencies. This is an important topic of study, since it appears that the beating of multiple NRP modes can trigger outbursts \citep{Rivinius2001, Rivinius2013}. There are, however, a multitude of other possible explanations for this observed periodicity.

Outbursts are generally understood as discrete events where material is transferred from the stellar surface to the inner region of the circumstellar disk, where it is then governed by gravity and viscosity. In visible photometric observations, outbursts are typically characterized by a `sudden' change in flux of the system, followed by a more gradual (relative to the initial change) decay back to baseline. Disk emission and polarization rise steeply during an outburst event \citep{Rivinius2013}, signifying an increasing density in the innermost regions of the disk, but the net change in brightness depends on the inclination angle of the system and can be positive or negative. If viewed edge-on or very nearly edge-on, the disk will obscure the central star and the system will appear dimmer as the disk grows, but at low to moderate inclination angles (e.g. $\sim$pole-on) the system will appear brighter with disk growth \citep{Sigut2013}. Models described in \citet{Sigut2013} where a disk grows over the course of a year and then dissipates over the course of two years show changes in V-band magnitude of a few tenths of a magnitude throughout this process, although this can vary depending on orientation, gravitational darkening due to rapid rotation, spectral sub-type, disk density, scale height, and the details of disk growth and evolution. \citet{Haubois2012} predict V-band brightening of up to 0.4 mag ($i$ = 0$^{\circ}$), and dimming of 0.2 mag ($i$ = 90$^{\circ}$) during disk growth/dissipation phases using the viscous decretion disk model. At an inclination angle of $i\sim$70$^{\circ}$, the net change in optical flux during an outburst is predicted to be nearly zero, because the additional emission and absorption introduced by the disk effectively cancel each other out.

Although outbursts are commonly seen in classical Be stars, their frequency, duration, and amplitude vary greatly from star to star, and a given Be star can show large variation in its outbursts over time. It is therefore necessary to amass a large number of observed outbursts for as many systems as possible, in order to better understand their systematic behavior and possible correlations with the underlying stellar properties.

Some Be stars retain their disks over many years or decades, and in these cases it is not uncommon for disks to posses density waves, which can then travel around the disk on timescales hundreds of times longer than that of Keplarian motion at a given radius from the central star, typically at a period on the order of $\sim$10 years \citep{Okazaki1991,Papaloizou1992}. In spectroscopic observations, these global oscillation modes in the disk cause variations in the ratio of the violet-to-red (V/R) peaks typical of Be star emission lines \citep{Rivinius2013}. When the high density material is approaching the observer, the violet (V) peak is enhanced, and when the high density part is moving away from the observer the red (R) peak is enhanced. Density waves can also produce photometric variations as a line-of-sight effect, which largely depends on the inclination angle of the system. 

Classical Be stars are a very heterogeneous class of objects, and a given Be star may show all, some, or none of the aforementioned types of variability over some observing baseline. To better understand Be stars as a class of objects, it is necessary to observe a large number of Be stars for a long time, with as many techniques as possible. This strategy will give a more complete and probabilistic picture of the characteristics of Be stars as a population. This is a necessary step, since results from a small number of well-understood Be stars can not accurately be extrapolated to the entire population. Any theoretical models describing Be stars and their disks must accommodate the severe inhomogeneity seen amongst Be stars as a population, and can not be specifically tailored to just a few well understood cases.

In this paper, we explore the variability of over 500 known Galactic Be stars, primarily through the use of time-series photometric data. In \S \ref{sec:data} we introduce the data products that make up the core of our analysis. \S \ref{sec:analysis} is a discussion of the various methods used to analyze the data, describing how each type of signal is recovered from a KELT light curve. In \S \ref{sec:results} we present our results. General patterns and trends are explored, and specific interesting cases are highlighted. In some cases, archival time-series spectroscopic data are included in our analysis, providing a more comprehensive view of such systems. In \S \ref{sec:conclusion}, we summarize the ensemble properties of our sample and discuss some highlights. Finally, the appendix includes plots for every variable object in our sample, as well as a brief discussion of each eclipsing system that we found.

\section{Data}
\label{sec:data}
\subsection{Kilodegree Extremely Little Telescope (KELT)} \label{KELT_intro}
The Kilodegree Extremely Little Telescope (KELT) is a photometric survey using two small-aperture (42 mm) wide-field (26$^{\circ}$ x 26$^{\circ}$) telescopes, with a northern location at Winer Observatory in Arizona in the United States, and a southern location at the South African Astronomical Observatory near Sutherland, South Africa. The KELT survey covers over 60$\%$ of the sky and is designed to detect transiting exoplanets around stars in the magnitude range $8 < V < 10$, but obtains photometry for stars between $7 < V < 13$. Designed for high photometric precision of better than 1$\%$, KELT's observing strategy involves long baselines of up to 10 years, with light curves for $\sim$4.4 million objects at the time of this writing (September 2016). KELT's effective bandpass is equivalent to a broad R-band filter. The long baseline combined with a typical cadence of 30 minutes and high photometric precision makes the KELT dataset a valuable resource for studying variable stars across a range of timescales and magnitudes. KELT uses a German Equatorial Mount, requiring data acquired in the eastern orientation and western orientation to be reduced separately \citep{Pepper2007,Pepper2012}.  For most objects observed by KELT, there are both raw and detrended versions of the light curves available. The detrending process is built into the KELT pipeline, and uses the Trend Filtering Algorithm (TFA) \citep{Kovacs2005} as implemented in the $VARTOOLS$ package \citep{Hartman2012}. The nearest 150 stars within two instrumental magnitudes of the target star (outside of a 20 pixel exclusion zone centered on the target star) provide the photometric reference for the detrending. Additionally, outliers are removed and long-term trends are subtracted out.

The per-point photometric errors for KELT observations are small, typically a few mmag for brighter sources, and up to a few percent for the faintest targets. The typical photomtric error for the KELT data in this analysis is 7 mmag. We find that our photometric errors are dominated by a complex combination of systematic noise sources (which are generally larger than our photon errors), so the typical errors quoted above are based on the empirical scatter of KELT light curves that do not display obvious variability. With each light curve having thousands of data points, plots can quickly become very cluttered with the inclusion of error bars. Or, depending on the scaling of the figure, the sizes of the error bars can be comparable to or smaller than the plotted data points themselves. For these reasons, we choose not to display error bars in any light curve plots. 

The KELT light curves are a good tool by which to study the photometric variability of Be stars, which can show variability on timescales ranging from hours to decades. KELT's bandpass primarily probes the stellar photosphere and the inner-most region of Be disks, out to a few stellar radii \citep{Carciofi2011}. The longest timescales of variability occurring over the course of months to many years, such as those owing to variations in the size and density of the circumstellar disk, are captured by the long baseline of observations. Shorter timescales of days to weeks are accessible for both periodic (e.g. binarity induced) and non-periodic (e.g. outbursts) events thanks to a high cadence (typically $\sim$30 minutes) and nightly observations. Variability at even shorter timescales of hours to a few days are accessible for periodic signals, such as those caused by pulsation, which can be recovered through the use of algorithms designed to find periodic signals in unevenly sampled time-series photometry.

\subsection{Be Star Spectra (BeSS) Database}

The Be Star Spectra (BeSS) database is a continually updated catalog that attempts to include all known Be stars, as well as their stellar parameters. This catalog is based primarily on the catalog of classical Be stars published by \citet{Jaschek1982}, but also includes more recently discovered Be stars from a variety of sources \citep[e.g.][]{Neiner2005,Martayan2006}. The BeSS database is updated regularly as new Be stars are discovered, confirmed, or dismissed in the literature \citep{Neiner2011}. At the time of this writing (September 2016), there are 2,320 Be stars in the BeSS database, with 118,657 total spectra available for 1,082 unique stars in the catalog. 

In addition to being the most up to date catalog of known Be stars, the BeSS database also contains spectra for about half of the included stars, with observations spanning a wide range of epochs and wavelengths, making it an extremely useful tool for studying Be star variability. The large number of objects and observations make the BeSS database valuable for statistical work in the study of Be stars. BeSS is updated daily to include new spectra submissions, but only after the BeSS administrators have validated each individual submission \citep{Neiner2011}. The contributed spectra are obtained by both professional and amateur astronomers, with many different telescopes and instruments. However, all data is submitted in the same format and undergoes the same scientific validity checks, and is considered to be of comparable quality. Most submitted spectra are taken in visible light, where evidence of the Be phenomenon is particularly apparent (the H$\alpha$ line being a good indicator of emission or shell absorption), but there is no restriction on the wavelength regime. Spectra are available for many of the Be stars under consideration in this paper, providing a valuable complement to the photometric data, especially in cases where there are time-series spectroscopic measurements simultaneous with the KELT light curve.

\begin{table*}
 \centering
 \caption{KELT fields containing B\MakeLowercase{e} stars}
 \label{tbl:KELT_fields}
 \begin{tabular}{cccccccccc}
    \hline
    \hline
Field & $\#$ of BK & Observing & $\#$ of Images & Date of First & Date of Last & Field Center & Field Center & Field Center\\
      &  Stars & Seasons & Processed & Observation & Observation & (RA; deg) & (Dec; deg) & ($l$,$b$)\\
    \hline
N01 & 2   & 6  & 3383  & 2006-10-26  & 2014-12-25  & 1.50 & +31.67  & (111.7, -30.2) \\
N03 & 11  & 7  & 8610  & 2006-10-25  & 2013-03-12  & 59.54  & +31.67 & (163.1, -16.3)  \\
N04 & 56  & 9  & 9721  & 2006-10-26  & 2014-12-31  & 88.56  & +31.67 & (178.6, 3.0) \\
N10 & 5   & 7* & 7550  & 2007-01-28  & 2015-07-17  & 262.68  & +31.67 & (55.6, 30.0) \\
N11 & 50  & 7* & 6011  & 2007-05-29  & 2013-06-13  & 291.70  & +31.67 & (65.0, 7.0) \\
N12 & 15  & 7* & 5159  & 2007-06-08  & 2013-06-14  & 320.72  & +31.67 & (79.3, -13.0) \\
N13 & 3   & 6* & 4767  & 2006-10-26  & 2012-06-22  & 349.74  & +31.67 & (100.9, -27.2) \\
N16 & 105 & 3  & 3383  & 2012-05-21  & 2014-12-29  & 0.80  & +31.67 & (111.1, -30.0) \\ 
N17 & 127 & 3  & 3268  & 2012-09-17  & 2014-12-29  & 40.80  & +57.00 & (137.7, -2.6) \\ 
J06 & 30  & 4  & 5184  & 2010-03-02  & 2015-05-06  & 114.90  & +3.00  & (215.8, 12.1) \\ 
S05 & 47  & 4  & 2893  & 2010-02-28  & 2015-11-19  & 91.80  & +3.00  & (205.2, -8.43) \\ 
S13 & 31  & 5  & 3566  & 2010-03-19  & 2015-10-05  & 276.00  & +3.00  & (32.5, 7.5) \\ 
S14 & 20  & 5  & 3384  & 2010-04-11  & 2015-10-25  & 299.00  & +3.00  & (43.2, -12.9) \\ 
S18 & 1   & 4  & 5568  & 2010-06-30  & 2015-11-19  & 23.00  & -53.00  &  (289.4, -63.1) \\ 
S19 & 1   & 3  & 3993  & 2010-02-28  & 2015-11-19  & 46.00  & -53.00  & (268.6, -54.4) \\ 
S20 & 1   & 3  & 4243  & 2010-02-28  & 2015-04-03  & 69.00  & -53.00  & (261.1, -41.5) \\ 
S21 & 3   & 5  & 5740  & 2010-02-28  & 2015-05-06  & 91.80  & -53.00  & (261.0, -27.8) \\ 
S22 & 11  & 4  & 4678  & 2010-03-02  & 2015-11-19  & 138.00  & -20.00  & (248.6, 19.0) \\ 
S23 & 2   & 4  & 4782  & 2010-03-12  & 2015-07-05  & 161.00  & -20.00  & (266.4, 33.7) \\ 
S24 & 1   & 4  & 5002  & 2010-03-12  & 2015-08-03  & 184.00  & -30.00 & (293.9, 32.2)  \\ 
S25 & 4   & 4  & 5593  & 2010-03-12  & 2015-08-15  & 207.00  & -30.00  & (317.3, 31.3) \\ 
S27 & 2   & 4  & 4990  & 2010-03-18  & 2015-11-12  & 299.00  & -53.00  & (345.3, -30.9) \\ 
S34 & 82  & 3  & 8416  & 2010-01-03  & 2015-11-19  & 123.40  & -54.00  & (269.0, -10.7) \\ 

   \hline
    \hline
 \end{tabular}
 \begin{flushleft}
  \footnotesize \textbf{\textsc{NOTES}} \\
  \footnotesize Observing seasons marked * are not well defined and have gaps due to the Monsoon season. Field S34 includes commissioning data. Field J06 is observed by both KELT-North and KELT-South.
  \end{flushleft}  
\end{table*}  

\subsection{KELT light curves of stars in BeSS}
From the BeSS database, we compiled a list of all the classical Be stars with $7 < V < 13$, which included 1362 unique objects. This magnitude range was chosen to align with KELT's magnitude limits. This subset was then cross-matched with the KELT catalog, and 610 unique objects that exist in both datasets were recovered. Of the 610 stars in this sample, 374 are observed by KELT North, 206 are observed by KELT South, and 30 are observed by both KELT North and South (the joint field J06). However, 100 of these light curves have saturation effects that can greatly complicate the analysis. This leaves 510 Be stars with KELT light curves that can be analyzed at present. There is no hard magnitude cut-off for an object being saturated in KELT, since the saturation threshold depends on the object's color, position on the chip, and other factors. Table~\ref{tbl:KELT_fields} contains information about each KELT field that is used in this study, including the number of known Be stars with KELT light curves, the number of images processed (which corresponds to the number of data points in the resulting light curves from that field), the span of observation dates, and the location of the field on the sky. 

Available in full in the supplementary material, Table~\ref{tbl:variable_class} in the text shows a sample of the BeSS stars with KELT light curves that are the subject of study in this paper. Unique, short identifiers are assigned to each object for convenience (the ``BK number''), beginning at BK-000 and incrementing up to BK-609. Celestial coordinates, identifiers, spectral types, and V-band magnitudes are gathered from the BeSS database. Periods and variable types, as determined from the KELT light curves, are also given here. 

\begin{table*}
 \centering
 \caption{Table of BeSS-KELT stars with variable type classifications}
 \label{tbl:variable_class}
 \begin{tabular}{ccccccccccc}
    \hline
    \hline
BK     & RA        & Dec      & V-mag & Spectral & Identifier & Period 1   & Period 2   & Period 3 &  Variable       \\
num    & (dd)      & (dd)     &       &   Type   &            & (days)     &  (days)    & (days)   & Type(s)         \\
    \hline
032    & 88.4994   & 26.4225  & 8.25  & B2Ve     &  HD 39478  &  0.54780   & ...        & ...      &  LTV,NRP        \\    
033    & 93.0578   & 20.0009  & 10.76 & Be       &  HD 253215 &    ...     & ...        & ...      &  LTV            \\    
034    & 81.3243   & 29.6149  & 8.93  & B1Ve     &  HD 35347  &  0.65516   & ...        & ...      &  LTV,NRP,ObV    \\    
035    & 81.4366   & 35.6472  & 8.43  & B1Vpe    &  HD 35345  & 28.19131   & ...        & ...      &  IP,LTV,ObV,SRO \\    
036    & 87.2235   & 29.1361  & 8.05  & B3pshe   &  V438 Aur  & 24.18518   & ...        & ...      &  IP,LTV,ObV     \\    
037    & 91.7265   & 19.0345  & 9.34  & B1Ve     &  HD 251726 &  1.35791   & 14.640840  & ...      &  IP,NRP,ObV     \\    
   \hline
    \hline
 \end{tabular}
 \begin{flushleft}
  \footnotesize \textbf{\textsc{NOTES}} \\
  \footnotesize 
  
  This table is published in its entirety in the machine-readable format. A portion is shown here for guidance regarding its form and content. In addition to the columns included here, the machine-readable version includes the following: number of archival spectra in BeSS, $v\sin{i}$ from BeSS, KELT identification numbers, KELT instrumental magnitudes, galactic longitude and latitude, 2MASS J, H, and K magnitudes \citep{Skrutskie2006}, WISE W1, W2, W3 and W4 magnitudes \citep{Wright2010}, and GALEX NUV and FUV magnitudes \citep{Martin2005}. Variable types are as follows. ``ObV'': Outburst Variation - outbursts are present in the raw light curve; ``SRO'': Semi-Regular Outbursts - outbursts occur with some regularity; ``LTV'': Long Term Variation - long term variability in the raw light curve; ``NRP'': Non-Radial Pulsator candidate - shows periodic variability at timescales of less than 2 days; ``IP'': Intermediate Periodicity - shows periodic variability at timescales greater than 2 days; ``EB'': Eclipsing Binary; ``DW(S/I)'': Double Wave - indicates double-waved modulation at (S)hort or (I)ntermediate periods; ``SAT'': Saturated - saturation issues in the KELT photometry make analysis of this star intractable at the present time. Variable types followed by ``?'' indicate some uncertainty in the ascribed characteristic.
  \end{flushleft}  
\end{table*}

\section{Time-series Analysis of KELT Photometry}
\label{sec:analysis}
We employ several techniques to identify different types of photometric variability (both aperiodic and periodic) in the light curves of the BeSS-KELT sample. We then interpret the behavior in terms of the possible physical mechanism(s) responsible for the observed variability, which are summarized in Table~\ref{tbl:variable_class}, and are explained in more detail in this and the following section.

It is often useful to consider separately the spectral sub-types of the systems under consideration. In what follows, we define ``early-type" Be stars as those with spectral types earlier than B4, ``mid-type" Be stars have spectral types including B4, B5 and B6, and ``late-type" Be stars have spectral types including B7 and later. Stars without a specific spectral type listed (e.g. a spectral type of ``Be'') in BeSS are considered ``unclassified''.

Because of the large pixel scale of KELT (23'') and the relatively crowded fields lying in or near the galactic plane where most Be stars are found, light from other sources is often blended with the target star in KELT. If a neighboring star is blended with the target star and is variable, then this variability can appear in the light curve for the target. This contamination is addressed by analyzing difference images of the pixels in the vicinity of the target to determine precisely which pixels are the source of variability. This analysis was done for all stars showing any type of photometric variability. This process can robustly identify contaminating sources further than two KELT pixels away from the target. Another consequence of blending is that the amplitude of variability in the target star will be diluted depending on how much flux from neighboring sources leaks into the target star's aperture.

Identifying aperiodic features requires careful visual examination of the light curve of each object in the sample. We attempted to use a few variability statistics \citep[e.g. $\Delta m$-$\Delta t$ plots and peak-finding;][]{Findeisen2015} in order to identify and quantify the different types of variability present in the sample, but the features seen are so disparate and varied that none of the statistical tests we applied were satisfactory, and a `by-eye' analysis was deemed to be more reliable. A number of qualitatively different features are seen, including sudden dimming/brightening events followed by a return to baseline brightness, seemingly random (and sometimes persistent) `flickering' at different timescales and amplitudes, gradual dimming/brightening events that take place over years and may or may not return to a baseline flux, and slowly oscillating (but not necessarily strictly periodic) brightness over the course of months or years. Multiple flavors of variability are often seen in the light curve for a single star, which can greatly complicate the analysis. 

\subsection{Identifying Long-Term Variation}

The long baseline of KELT light curves allows for the detection of long-term variability (LTV) on the order of years. Variability at these timescales is generally attributed to changes in the circumstellar disk including growth, dissipation, or periodic density oscillations \citep{Haubois2012}. The raw KELT light curves for all stars in the sample were visually inspected for signs of LTV. Some systems remain photometrically stable for years, then begin to gradually dim or brighten at a rate of a few hundredths of a magnitude per year. Others seem to oscillate around an average brightness, while others show gradual variability interspersed with outbursts of relatively higher amplitude. We classify all such stars as belonging to the LTV category, so long as we have a photometric baseline of at least four years.

\subsection{Identifying Outburst Variation}
Because Be star outbursts show a large range in their amplitude, duration, and morphology, they are difficult to rigorously define. In what follows, a feature in a light curve is considered an outburst if there is a sharp departure from baseline (either brightening or fading) followed by a gradual return towards baseline \citep{Sigut2013}. Despite this simple description, there are numerous difficulties involved with detecting outbursts. In many cases, there is no well-defined baseline, and outbursts are superimposed on stochastic or long-term variability. A Be star may experience outbursts, but at amplitudes below KELT's detection threshold. \citet{Balona2015} use Kepler data to find outbursts with amplitudes of $\lesssim$ 10 mmag and durations of days in the Be star KIC 6954726. Such low-amplitude and aperiodic features are not resolvable in KELT light curves. Furthermore, we expect no appreciable net change in flux during outbursts for systems with i$\sim$70$^\circ$, as explained in the introduction.

In order to count the number of outbursts seen in the KELT light curve for a given star, each season of observation was visually inspected for outburst signatures, and the number of outbursts tallied. Many situations arise that make counting the number of outbursts convoluted. For example, a star may brighten suddenly and begin to decay back to baseline, but then suddenly brightens again shortly after the first brightening. This example would count as two outbursts, even though the first outburst never fully decayed back to the pre-outburst brightness. The outburst rates for each star were calculated by adding up the length (in days) of each observing season, converting this number to years, and then dividing the number of outbursts by the total number of years observed. This was chosen because the duration of a single outburst tends to be shorter than a single observing season. If instead the number of outbursts was divided by the full baseline of observation (including the $\sim$150 day gaps between seasons), then the outburst rates would be systematically underestimated for a majority of the stars. The few exceptions to this (i.e. cases where the chosen method overestimates outburst rates) are stars exhibiting very long outbursts lasting for multiple seasons. The outburst rates claimed for each star are an approximate lower limit, because the outbursts were judged by eye and are not mathematically defined, and outbursts with amplitudes below the detection threshold in a KELT light curve may be present. Future work may attempt to quantify outburst rates in a more rigorous way. 

\subsection{Identifying Periodic Signals}
Light curves for all Be stars in the BeSS-KELT sample were analyzed for periodic variability using a Lomb-Scargle Periodogram \citep[LSP;][]{Lomb1976,Scargle1982}. When searching for periodic behavior in Be stars, it is useful to make a distinction between `short' and `intermediate' variability. In what follows, short periodic variability corresponds to periods of less than two days, and intermediate periodic variability occurs on timescales between two and 200 days. This choice was made because typical pulsation modes in Be stars occur with periods of two days or less. Despite this distinction, short period variability must be probed simultaneously with intermediate timescale variability due to the aliasing of low-frequency modes with the diurnal sampling rate. For example, a star with a 10 day period will show aliases at 0.9091 and 1.1111 days from the combination of frequencies of 1 d$^{-1}$ (diurnal observing) and 0.1 d$^{-1}$ (astrophysical periodic variability). Without knowledge of the 10 day period, this example would be misidentified as a system showing short-period variability. Using algorithms to identify the LSP peaks with the highest power will very often select aliases of this nature, because the aforementioned aliases often have a LSP power comparable to or higher than a signal caused by true astrophysical variability. Therefore, the entire frequency range of the LSPs for all systems had to be visually inspected in order to mitigate the misidentification of aliases as real variability modes. 

When searching for high frequency signals (like those associated with NRP), a high-pass median filter, as well as 4 $\sigma$ clipping to remove outliers, was applied to each light curve, as implemented in the $VARTOOLS$ light curve analysis package \citep{Hartman2012}. This smooths out all long term variability, making it easier to identify relatively short periodic signals in the data. While performing this analysis, care was taken to mitigate false positives. Daily aliases associated with the diurnal observing strategy of KELT are common features in many of the the LSPs, occurring near integer fractions and multiples of one day. Systematics of this nature are inherent in any ground based survey, being especially pervasive in wide-field surveys like KELT. However, these systematics are fairly well understood in the context of the KELT data. Often the daily alias is the strongest LSP peak, so it is useful to create a whitened LSP, which filters out the strongest signal from the data and allows for the detection of relatively weaker signals. Whitened and non-whitened versions of a LSP were then generated for these median-smoothed light curves, which were phased to the top six peaks (in both whitened and non-whitened versions). A sinusoid was fit to each phased light curve, and its amplitude (peak-to-trough) and the median absolute deviation (MAD) of the residuals to the sinusoidal fit were calculated. The ratio of the signal amplitude to the MAD of the residuals was used to parametrize the strength of the signal relative to the scatter in the data, serving as a metric for the reliability of the recovered signal. Cases where this ratio is less than 0.75 are deemed to be non-detections. For each light curve with a signal giving a ratio higher than 0.75, the LSP and phased light curves were visually inspected to determine which, if any, is the best period for the system. The same analysis was done for the TFA detrended version of each light curve. To qualify as a detection, a LSP peak must exist at the same period for both the detrended and median-smoothed versions. In most cases, the period with the highest Lomb-Scargle power was chosen to be the best period for the system. Exceptions to this were made when the top LSP peak was very close to an integer fraction of a day (within 0.2\%), and when there was strong Lomb-Scargle power at other aliases associated with the diurnal sampling of KELT, indicating telescope systematics as the cause of the signal. These light curves were then pre-whitened to the top period, and the next strongest peak was inspected. 

Intermediate periodic variability was probed with LSP up to periods of 200 days. This was done simultaneously, but independently for the raw, TFA detrended, and median-smoothed versions of each light curve. In many cases, the detrended data was best suited for this periodicity search. The process of detrending removes the effects of telescope systematics and outliers, providing a cleaner light curve for the Lomb-Scargle analysis, resulting in a stronger Lomb-Scargle signal at higher confidence. However, for some systems it was more appropriate to use raw data. These were typically cases with longer periods, high amplitudes, and/or multiple outbursts. The detrending process includes outlier removal, which essentially removes most of the data taken during outburst phases. For a signal to count as a detection, it must exist in multiple versions of the light curve.

\section{Results} \label{sec:results}
While \S \ref{sec:analysis} is organized according to the photometric behavior being probed, this section organizes the results of that analysis according to the underlying physical phenomena to which we attribute the observed behavior. Characteristic examples are displayed in this section, and plots for every object showing variability are presented in the appendix. For ObV and LTV variables, the raw KELT light curve is displayed, while results for NRP, IP, SRO, and EB variables are displayed in a phased light curve.

\subsection{Outbursts} \label{sec:outbursts}
\begin{figure}[!ht]
\centering\epsfig{file=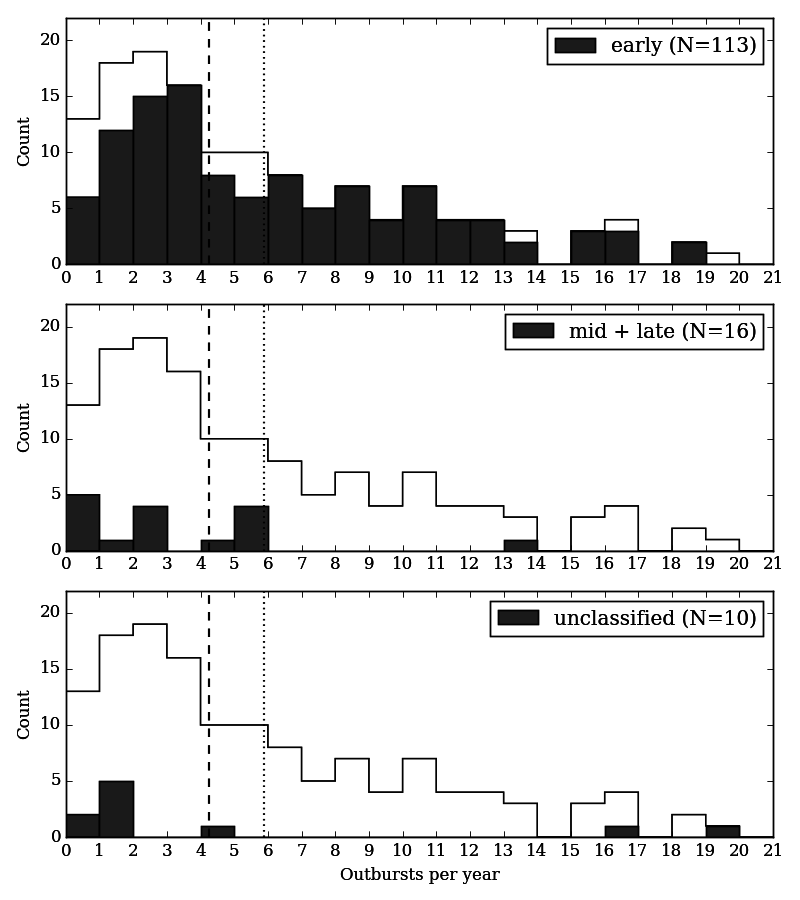,clip=,width=0.99\linewidth}
\caption{Distribution of outburst rates for early (\textit{top}), mid + late (\textit{middle}), and unclassified (\textit{bottom}) spectral types. These were calculated for all systems that show one or more outburst, so long as the number of outbursts is well defined. The solid line making up the envelope of the distribution in all three panels includes all stars, regardless of spectral type. The vertical dashed line denotes the median, and the vertical dotted line denotes the mean of this distribution as a whole.}
\label{fig:Outburst_rates_early}
\end{figure}

The outburst rates (number of outbursts per year) were calculated for all stars where they could be reliably determined. In cases where the number of outbursts was ambiguous or otherwise uncertain, the star was not included in the final statistics. In total, there are 140 stars in this sample where the number of outbursts could be determined. The outburst rates for these are shown in Figure~\ref{fig:Outburst_rates_early}. It is apparent that there are a wide range of outburst rates in this sample, but it must be noted that neither the amplitude nor the duration of an outbursts is considered in this distribution.
We find that 36$\%$ of the BeSS-KELT sample exhibits one or more outburst in its KELT light curve, with a higher occurrence rate (51$\%$) seen in early-type stars compared to mid- (20$\%$) and late- (5$\%$) types. These fractions were calculated from the subset of the sample where it was unambiguous whether or not outbursts were present. There is a wide range in the number of detectable outbursts in the light curves of the stars in our sample, with some Be stars showing zero or one outburst, while others undergo several dozen outbursts throughout the observing baseline.

\begin{figure}[!ht]
\centering\epsfig{file=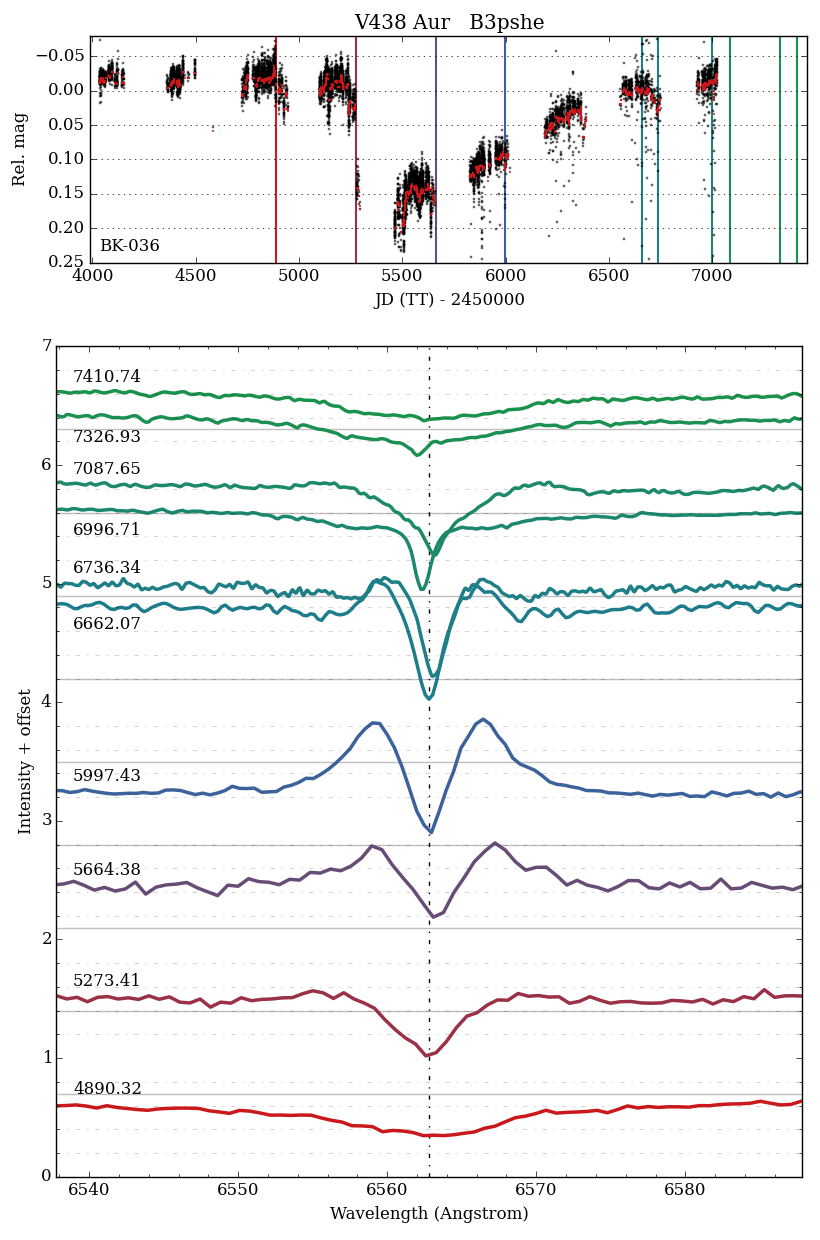,clip=,width=0.99\linewidth}
\caption{Photometric outburst that is correlated with disk creation. This shell Be star (HD 38708; B3pshe) experiences a large outburst beginning at around JD$_{\rm TT}$ = 2455200. {\it Top:} KELT light curve, with vertical lines corresponding to dates of spectroscopic observations from BeSS. The black points are the raw data, and red points show a version of the light curve with a low-pass filter applied. {\it Bottom:} H$\alpha$ line for each of these spectra, increasing in time from bottom to top, and spaced according to the JD$_{\rm TT}$ - 2450000 epoch of observation, which is listed on the left.}
\label{fig:Outburst_disk}
\end{figure}

To examine the idea that photometric outbursts correspond to material being transferred from the stellar surface to a circumstellar disk, we look for cases where a star has spectroscopic measurements both before and after a photometric outburst. If outbursts do indicate mass transfer from star to disk, then we would expect to see evidence for this in the spectra, where the signature of a circumstellar disk can be inferred. A KELT light curve for the Be star BK-036 (HD 38708), showing a clear outburst in photometry is displayed in Figure~\ref{fig:Outburst_disk}. Ten spectra spanning $\sim$7 years from BeSS provide good spectroscopic coverage of the outburst. During the pre-outburst phase (prior to JD$_{\rm TT}$ = 2455200) the light curve is flat and the H$\alpha$ line shows no evidence of a disk. The second H$\alpha$ measurement occurs during the onset of the outburst, and shows deeper shell absorption with emission wings beginning to emerge. The system then reaches its dimmest point (JD$_{\rm TT}$ $\sim$ 2455500), when the recently ejected material obscures the stellar photosphere the most in KELT's bandpass. The disk then dissipates, which we see as a gradual return to the system's pre-outburst flux level. The H$\alpha$ line continuously evolves throughout this process, first developing large emission wings, then moving towards more narrow shell-absorption. In the final spectroscopic measurement, the H$\alpha$ line has returned to its pre-outburst line profile and shows no evidence of a disk. Examining systems like HD 38708 where a photometric outburst is bracketed by spectroscopic measurements provides strong evidence linking photometric outbursts with the creation of a circumstellar disk.

\begin{figure}[!ht]
\centering\epsfig{file=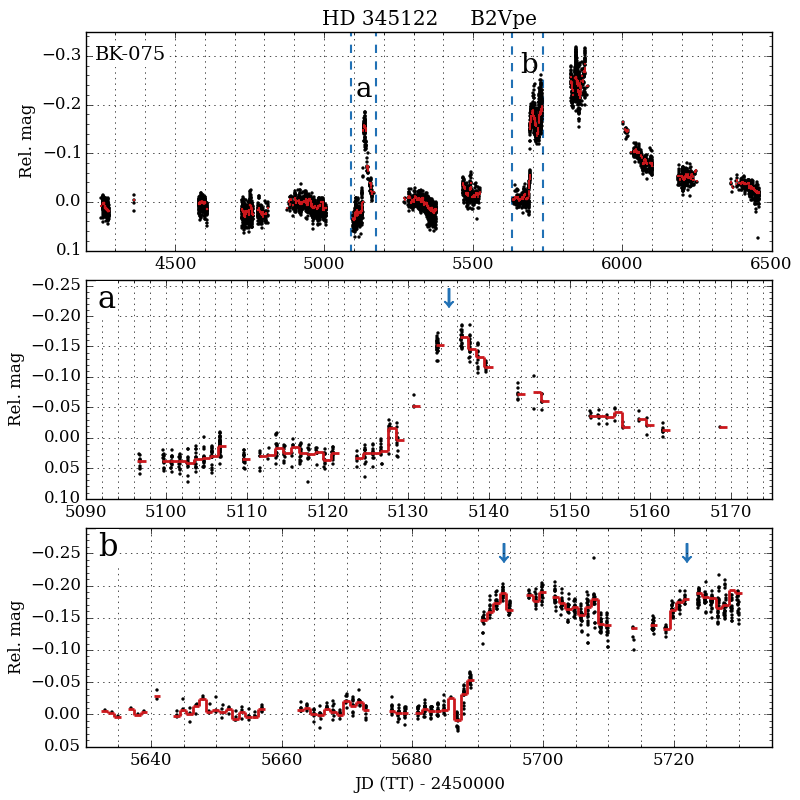,clip=,width=0.99\linewidth}
\caption{{\it Top:} Raw KELT light curve for BK-075 (HD 345122; B2Vpe). A relatively short outburst occurs near JD$_{\rm TT}$ = 2455100 (outburst `a'), followed by a quick return back to baseline. Around 560 days later there is another outburst (outburst `b') that is larger in amplitude and much longer in duration. {\it Middle:} A more detailed look at the region marked by the two vertical dashed lines bracketing outburst `a' in the upper panel. This highlights the features of the first outburst, with an arrow indicating the time of maximum brightness. {\it Bottom:} A zoom-in on the rising phase of outburst `b', showing more complexity than outburst `a', with arrows marking the two distinct brightening events that are separated by about 28 days. The red line shows the data median-binned with a bin size of one day in both lower panels.}
\label{fig:outbursts_075}
\end{figure}

\begin{figure}[!ht]
\centering\epsfig{file=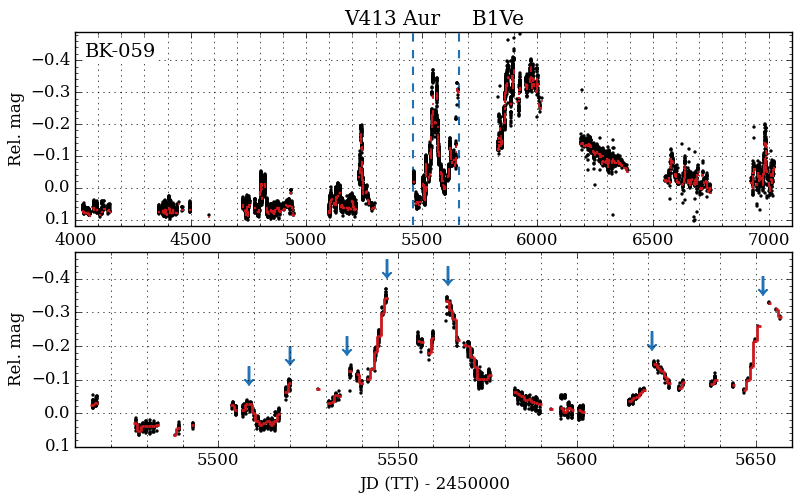,clip=,width=0.99\linewidth}
\caption{{\it Top:} Raw KELT light curve for BK-059 (HD 33152; B1Ve), a Be star with many irregularly spaced outbursts of varying amplitude and duration, together with longer term variability. {\it Bottom:} A more detailed look at the region marked by the two vertical dashed lines in the upper panel, with arrows indicating discrete outburst events. The red line in the lower panel shows the data median binned with a bin size of one day.}
\label{fig:outbursts_059}
\end{figure}

Not all photometric signatures of outbursts are as clear as the case of HD 38708. Figure~\ref{fig:outbursts_075} shows the KELT light curve for BK-075 (HD 345122, upper panel), classified here as an ObV variable. The first outburst is enlarged in the middle panel of the same figure, and exhibits a characteristic quick rise and slow decay indicative of a typical, isolated outburst. The larger (both in amplitude and duration) outburst that follows (near JD$_{\rm TT}$ $=$ 2455600) is actually comprised of at least two discrete outbursts which build on each other before the system starts decaying back to baseline. The rise of this double-outburst is displayed in the lower panel of Figure~\ref{fig:outbursts_075}, where it is clear that this is more complex than the prior outburst. We identify four outbursts in HD 345122, for an outburst rate of 1.7 outbursts/year. An even more convoluted example is shown in Figure~\ref{fig:outbursts_059} for BK-059 (HD 33152). This system shows numerous outbursts that are irregularly spaced and of varying amplitude and duration, coincident with slower brightness variation on timescales of years. The net brightening of the system leading up to JD$_{\rm TT}\sim$2456000 can be explained by a growing circumstellar disk that accumulates material following each outburst event, and is replenished (by outbursts) faster than it is dissipating. As the outbursts become less frequent and/or weaker, the disk dissipates faster than it is being replenished, and the system begins to return to its baseline brightness. BeSS spectra for this object show a single-peaked H$\alpha$ line in emission, indicating that this system is oriented $\sim$pole-on ($i\lnsim45^{\circ}$). The lower panel of Figure~\ref{fig:outbursts_059} shows a more detailed view of a single season of KELT data for this object, with individual outbursts marked with arrows. We identify 37 outbursts in HD 33152, for an outburst rate of 8.8 outbursts/year. These three systems highlight the diversity of outbursts seen in the light curves of Be stars.

Among the 168 Be stars in our sample with one or more outburst, we find that 35 (21$\%$) of these have outbursts occurring with enough regularity to be considered semi-regular outburst (SRO) variables. In most cases, these outbursts are not strictly periodic, but tend to occur at a roughly constant frequency, with varying amplitudes and some cycles being skipped, and/or occasional outbursts occurring between cycles. There are also cases where outbursts are roughly periodic within a certain segment of the light curve, but are not regular throughout the entire baseline of observation. 

Broadly speaking, there are two explanations invoked to explain SROs. The first is that there is some internal mechanism responsible for modulating the outburst behavior. This internal mechanism may be stable, or it may turn on and off over time. The second possibility involves a Be star in an eccentric binary system. At periastron passage, the non-Be star component of the binary may exert enough of a gravitational influence to trigger an outburst in the Be star if conditions are right. The characteristics of the resultant outburst would then depend on a combination of the gravitational triggering and the internal conditions of the Be star. Since vibrational modes in some Be stars are observed to change in amplitude and frequency, we would expect the combination of regular periastron passage and irregular internal conditions to result in semi-irregular (in amplitude and timing) outburst events in this scenario. The Be star $\delta$ Sco is known to be in a highly eccentric binary system with an orbital period of $\sim$10.6 years. \citet{Miroshnichenko2001} analyzed spectra during the system's periastron passage in 2000, and suggest a hypothesis where the NRP in the Be star component are amplified at periastron, triggering an increase in the mass loss rate. This system seems somewhat more complicated though, since spectroscopic data show that the disk began to form slightly before periastron \citep{Miroshnichenko2013}. However, the important conclusion that the mass loss rate is enhanced near periastron remains intact.  On the other hand, SRO have also been observed in Be stars ($\lambda$ Eri and $\mu$ Cen) that have no detected binary companions \citep{Mennickent1998,Rivinius1998}, and thus the SRO are presumed to be modulated by internal mechanisms. 

\begin{figure}[!ht]
\centering\epsfig{file=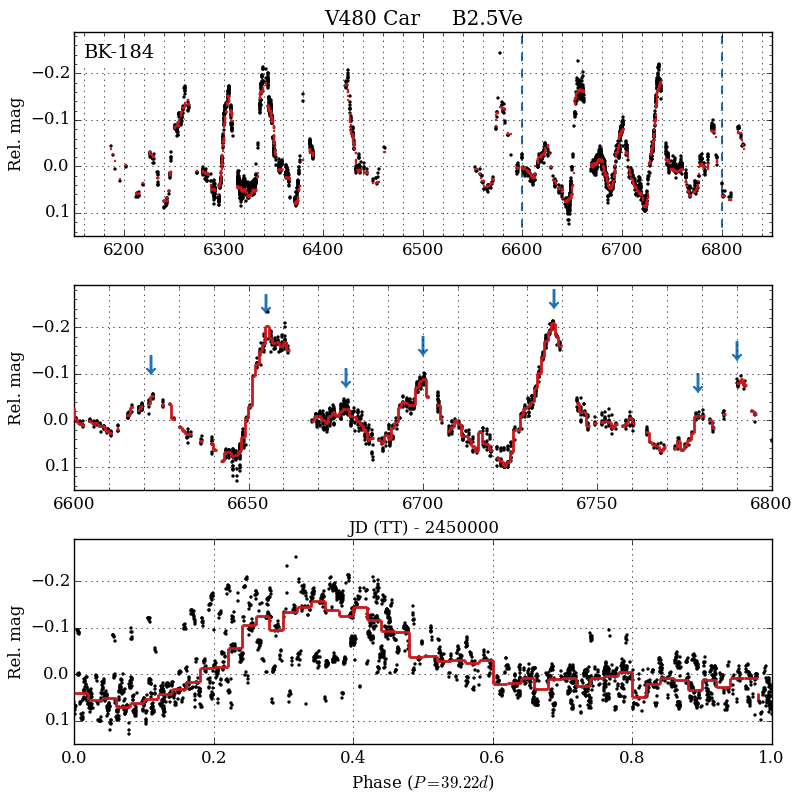,clip=,width=0.99\linewidth}
\caption{{\it Top:} Raw KELT light curve for BK-184 (HD 81654; B2.5Ve), a Be star with outbursts occurring semi-regularly every $\sim$40 days. {\it Middle:} A more detailed look at the region marked by the two vertical dashed lines in the upper panel. {\it Bottom:} The light curve is phased to a period of 39.22 days, highlighting the semi-regular nature of the outbursts. The outbursts do not align perfectly, but there is some degree of coherence in their occurrence. The red line in the middle panel shows the data median binned with a bin size of one day, while the lower panel uses a bin size of 0.02 in phase.}
\label{fig:outbursts_184}
\end{figure}

The star BK-184 (HD 81654) is an example of a system with SRO and a very high duty cycle, and is shown in Figure~\ref{fig:outbursts_184}. Including the commissioning data (not shown in Figure~\ref{fig:outbursts_184}), we identify 21 outbursts, and an outburst rate of 13.4 outbursts/year. It is clear from the raw light curve (upper panel) that the outbursts are not purely periodic, and also have varying amplitudes. However, when the light curve is phased to a period of 39.22 days (bottom panel), many of the outbursts are roughly aligned and binning the data traces what could be considered an `average' outburst for this star. \citet{Lefevre2009} use Hipparcos photometry to study variability among OB stars in a study that includes HD 81654. The authors list this star as having a variable type of `GCAS?' (indicating the variability is irregular and the variability type cannot be easily classified), a period of 40.036 days, and an amplitude of 0.228 mag. The Hipparcos mission operated between 1990 and 1993, and the KELT data shown here was collected between 2012 - 2014. The fact that virtually the same period (to within 2$\%$) is found in data taken $\sim$20 years apart indicates that the regularity of these outbursts is stable over decades. If this star is not part of a binary system and is in fact modulated by internal mechanisms (similar to $\lambda$ Eri and $\mu$ Cen), then this points towards the ``internal clock'' being remarkably stable despite the obvious variability seen in the KELT light curve. If, on the other hand, HD 81654 is a binary system, then it could prove to be an interesting case to study the role of binary interactions in triggering outbursts. Either way, this star is a good candidate for a more detailed investigation regarding the mechanism(s) responsible for the regularity of its outbursts. We do not detect any high frequency modes (the signature of NRP) in the KELT data for this star, but the LSP is completely dominated by both the 40 day signal, and aliases of this signal with the diurnal sampling. Pre-whitening against this signal does not seem to improve the situation, nor does applying a high-pass filter to smooth out long term trends. Therefore, we can neither detect nor rule out the presence of NRP.  The single available BeSS spectra for this object clearly shows H$\alpha$ in emission.

\subsection{Slow disk growth/decay, and disk oscillations}
\begin{figure}[!ht]
\centering\epsfig{file=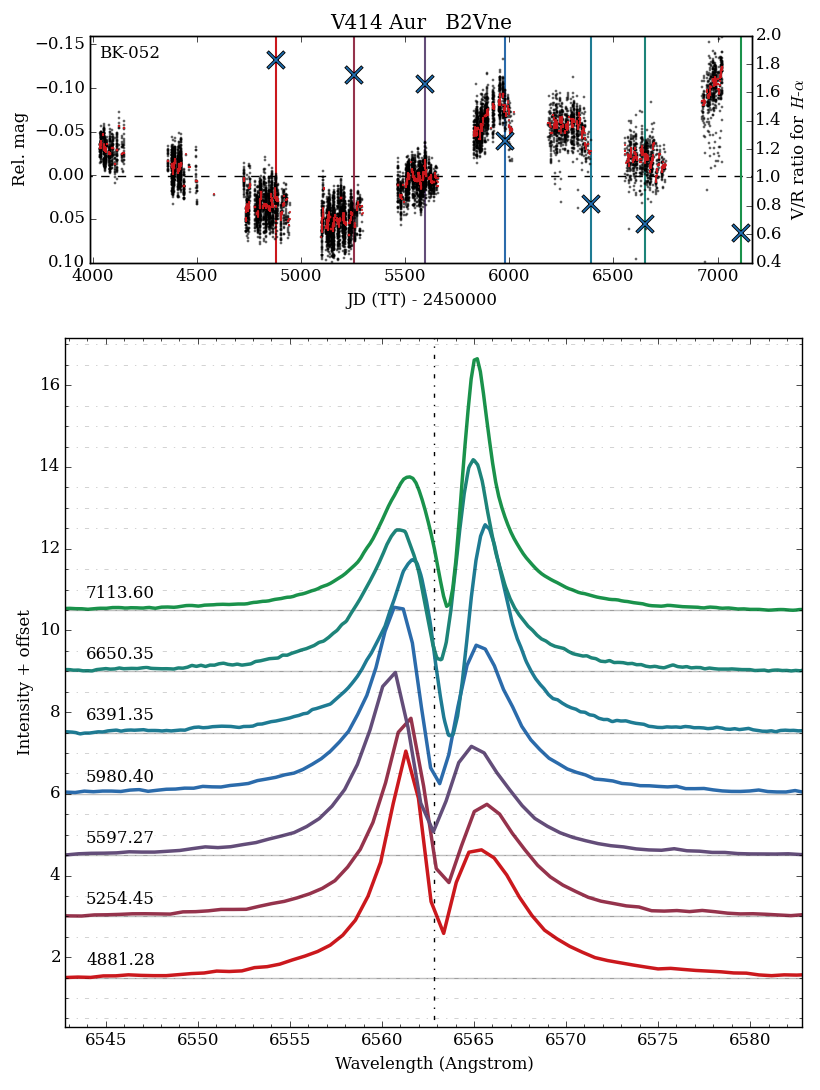,clip=,width=0.99\linewidth}
\caption{{\it Top:} Raw KELT light curve for BK-052 (HD 33232 = V414 Aur; B2Vne). The seven colored vertical lines correspond to dates of spectroscopic observations, and the blue X's show how the V/R ratio in the H$\alpha$ emission peaks is changing over time, as determined from the existing BeSS spectra for this target. {\it Bottom:} H$\alpha$ line profiles gathered from BeSS, increasing in time from bottom to top, show a gradual shifting in the V/R ratio over the $\sim$6 years of spectroscopic observation.  }
\label{fig:VR_var}
\end{figure}

From the 217 light curves that have a long enough baseline (4+ years) to detect LTV, we find that 80, or 37$\%$, of the BeSS-KELT sample belongs to this category. Splitting this by spectral sub-types, LTV is detected in 45$\%$ of early-type, 29$\%$ of mid-type, and 13$\%$ of late-type stars. An example of oscillatory LTV is shown in Figure~\ref{fig:VR_var}, for the star BK-052 (HD 33232). The upper panel shows that the light curve exhibits a coherent and roughly sinusoidal signal with a period of $\sim$7 years and an amplitude of $\sim$0.05 mag. This Be star also has simultaneous time-series spectroscopic measurements from BeSS which show variability at a similar timescale in the V/R ratio of the H$\alpha$ line (lower panel). These contemporaneous observations support the idea that the same mechanism (global density oscillations in the circumstellar disk) can be responsible for the long term variability seen in both the V/R ratio in the emission line profile and the coherent gradual changes in brightness of the system. While the H$\alpha$ line varies in its V/R ratio, the overall strength of H$\alpha$ emission is relatively constant, indicating that the disk as a whole is neither growing nor dissipating in a significant manner over the $\sim$6 years of spectroscopic observation. Be star disks do dissipate over time, so in order for the disk surrounding this star to exist in a quasi-steady state for six years there must occasionally be mass transferred to the disk. However, no outbursts are detected in the KELT light curve for this star. We suspect that this system does experience outbursts, but with amplitudes below the detection threshold of the KELT light curve. \citet{Merrill1952} analyze spectra for HD 33232 and trace the emission profile of the H-Beta line (amongst others) between 1943 - 1952, and find a similar oscillatory trend where the relative strength of the V and R peaks is variable. Although the theory of Be star disks was less developed at the time, their observations agree with the idea of a density wave moving around the disk. 

\citet{Haubois2012a} examine long-term photometric variability in the Be star 48 Lib, claiming correlations between its color and magnitude, and the long-term V/R variation. The authors comment that there is no indication the variability is caused by changes in the mass injection rate, but attribute the photometric variability to an azimuthal structure in the disk \citep[one-armed density wave;][]{Okazaki1991}. They also note that the H$\alpha$ equivalent width and emission height were constant. \citet{Mennickent1994} also monitor 48 Lib over a long baseline of $\sim$8 years, and observe long-term trends in both brightness and the V/R ratio. The authors note that there is a possible relation between intermediate brightness levels corresponding to extrema of V/R. Put another way, maximum and minimum brightness seems to occur when V/R is near unity. The same authors note the opposite trend in another Be star (V 1294 Aql), where brightness extrema correspond to V/R extrema. The case for HD 33232 appears different than both of these. The first photometric minima roughly corresponds to a V/R maximum, and when V/R is near unity the system is near a photometric maximum. Some combination of spiral-shaped density waves and different disk densities and inclination angles may be a reasonable explanation for the differing phase delays between photometric and V/R extrema in the three aforementioned systems.

\subsection{Non-radial pulsation}
\begin{figure}[!ht]
\centering\epsfig{file=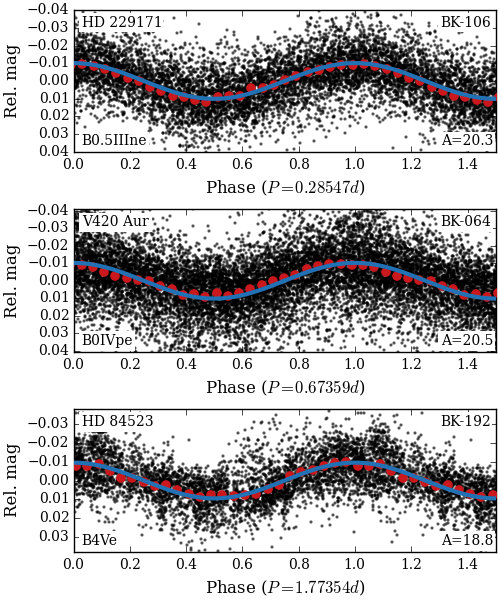,clip=,width=0.99\linewidth}
\caption{Light curves for three different stars phased to their recovered period. These phased light curves are typical of the signals we interpret as being caused by NRP. Red points show the light curve binned in phase using a bin size of 0.04, and the blue curve is a sinusoidal fit. The amplitude, given in the bottom-right corner of each panel, is the difference between the maximum and minimum points of the sinusoidal fit in units of mmag.}
\label{fig:NRP}
\end{figure}

\begin{figure}[!ht]
\centering\epsfig{file=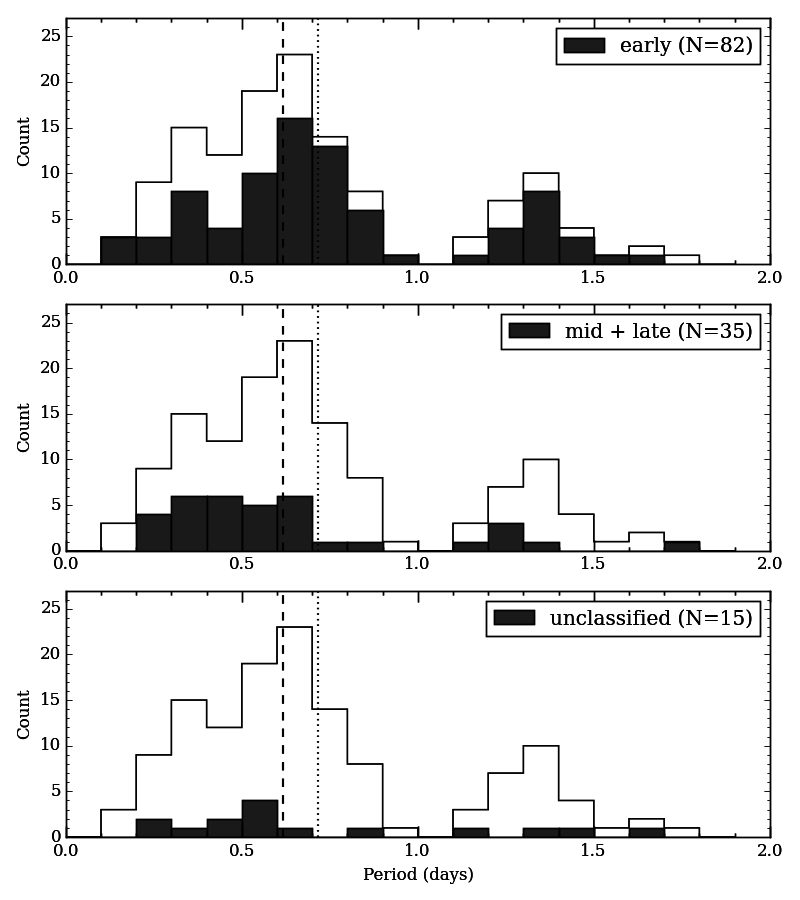,clip=,width=0.99\linewidth}
\caption{Histogram showing the distribution of high-frequency periodic signals, presented in the same manner as in Figure~\ref{fig:Outburst_rates_early}. The dearth of recovered periods close to one day is largely a consequence of the diurnal sampling of KELT, and should be interpreted as a systematic effect.}
\label{fig:hist_NRP_periods}
\end{figure}

We detect high frequency periodic variability, which we interpret as being indicative of NRP, in 25$\%$ of our sample. Incidence rates are highest for early-type Be stars (28$\%$), and lower for mid- (25$\%$) and late- (17$\%$) types. Figure~\ref{fig:NRP} shows examples of phased light curves for three such systems, and Figure~\ref{fig:hist_NRP_periods} shows the distribution of the recovered periods for all systems showing signs of NRP. It is important to note that the absence of a detectable periodic signal in a KELT light curve does not imply that the star is not pulsating, but rather suggests an upper limit to amplitudes of long lasting pulsational modes. Space based photometry has shown that pulsations in Be stars with amplitudes less than 1 mmag are common \citep{Gutierrez-Soto2008}, but this degree of precision is not realized in the ground-based KELT photometry. Additionally, Be stars are sometimes found to exhibit transient pulsational modes which last for a few days to months. These transient modes may precede an outburst, then disappear or diminish in amplitude following the outburst event \citep{Rivinius2013, Gutierrez-Soto2008, Huat2009}. Even if these transient modes are of a large enough amplitude to be detectable in KELT photometry, they are unlikely to be detected if they are present for only a small fraction of the total baseline of observation. In this analysis, we only look for periodic variability present throughout the whole baseline of observation. For these reasons, it is not surprising that we detect NRP in a significantly smaller fraction of our sample when compared to other studies that were specifically designed to detect these signatures, such as those mentioned in the introduction. 

\subsection{Other periodic variables (2 $\lesssim$ P $\lesssim$ 200 days)}
\begin{figure}[!ht]
\centering\epsfig{file=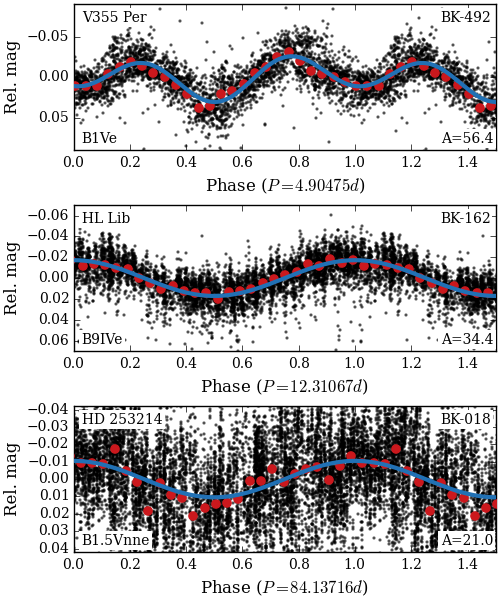,clip=,width=0.99\linewidth}
\caption{Same as Figure~\ref{fig:NRP}, but for IP variables. The top panel shows a double-waved signal, and is fit with a combination of two sinusoids.}
\label{fig:Intermediate_periodic}
\end{figure}

\begin{figure}[!ht]
\centering\epsfig{file=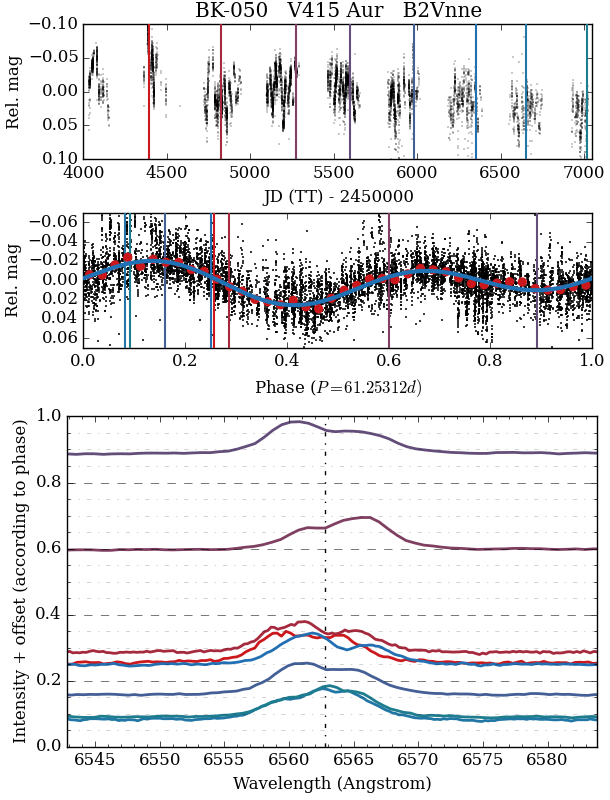,clip=,width=0.99\linewidth}
\caption{{\it Top:} Raw KELT light curve for BK-050 (HD 33461 = V415 Aur; B2Vnne), with vertical lines indicating dates of BeSS spectra. {\it Middle:} TFA detrended KELT data phased to a period of 61.25312 days. The time-series spectra are then phased to this period, with the color corresponding to the epoch of observation as indicated in the top panel. {\it Bottom:} Shown are the normalized H$\alpha$ profiles for each of the eight BeSS spectra offset by the photometric phase.
}
\label{fig:VR_var_phase_locked}
\end{figure}

In the BeSS-KELT sample, 40\% (201/510) of Be stars exhibit periodic behavior at intermediate timescales. Most of these are single-waved, and are well described by a single sinusoid. However, an appreciable fraction are double-waved, having unequal maxima and/or minima and requiring two or more sinusoids to describe their shape. Characteristic examples of these are shown in Figure~\ref{fig:Intermediate_periodic}, with the top panel showing a double-waved phased light curve, while the lower two are single-waved. With KELT photometry alone, we cannot further constrain the physical cause of this behavior like we do with the short period NRP candidates. Nonetheless, it is interesting to see such a large fraction of observed stars showing periodicity in this range. These objects are good candidates for continued investigation. One such example IP variable with a period of 61.25312 days and a clear double-wave modulation is shown in the upper panel of Figure~\ref{fig:VR_var_phase_locked}. This object (BK-050 = HD 33461) has eight BeSS spectra that are each spaced about a year apart and are simultaneous with the KELT light curve. When these spectra are phased to the photometric period, they show coherent variability. This seems to imply that the same mechanism is responsible for modulating both the brightness and spectroscopic line profile of the system. Cases like this, where spectroscopic data can be phased to a photometric period, can provide valuable clues for uncovering the underlying mechanism(s) causing the observed variability. 

There are numerous physical scenarios capable of giving rise to periodic variability longward of 2 days in Be stars. Isolated Be stars may experience the beating of NRP modes, Rossby modes, or circumstellar activity. As discussed in Section \ref{sec:outbursts}, the beating of multiple NRP modes can cause periodic variability, with the period depending on how closely spaced in frequency the modes are. Single vibrational modes may be modulated by the rotation of the star, resulting in Rossby modes with periods longer than those typically attributed to NRP \citep{Townsend2003}. Circumstellar activity owing to clumps of recently ejected material can cause observable variability. These clumps will have an orbital period that depends on their orbital radius, which may not be constant. Circumstellar processes of this nature are not expected to be strictly periodic. The shifting period and noise intrinsic in these processes will result in complicated frequency spectra, especially when considering aliases with both astrophysical and instrumental signals \citep{Baade2016}. Interactions between a Be star and a binary companion can also induce periodic variability modulated by the orbital period of the binary pair. Ellipsoidal precession of a Be star disk, tidally locked density waves in the disk, tidally induced disk warping, heating of the outer region of the disk by a hot companion, or the deformation of the stellar surface of one or both components can arise from gravitational interactions between the two binary components. Reflection effects may also be present. Global oscillations (i.e. density waves) in the circumstellar disk can also cause long-period variability in effectively single Be star systems, but these are considered separately, as the timescale is much longer (typically on the order of 10 years). Binarity is common amongst massive stars, and Be stars are not exceptions to this. We therefore expect an appreciable fraction of our sample to be in a binary system. \citet{Oudmaijer2010} use adaptive optics to probe the binary fractions of B and Be stars, with the sensitivity to detect binary companions separated by 20 - 1000 au. They find virtually the same binary fractions between B and Be stars (29 $\pm$ 8 $\%$ and 30 $\pm$ 8 $\%$, respectively), with similar underlying distributions of mass ratios, binary separations, and cumulative distributions. \citet{Moe2016} perform a meta-analysis of about 30 separate surveys, asserting that ``massive stars are dominated by interactions with binary companions''. We expect this statement to apply to Be stars since they are a subset of the massive star population.

Simulations for coplanar binaries including a Be star and disk show that in nearly all cases, the disk will exhibit periodicity in its structure that would otherwise be absent in an isolated system. In circular orbits the density structure will rotate with the orbital phase, and in elliptical orbits the disk brightness will change over an orbital cycle \citep{Panoglou2016}. These signals are potentially detectable in photometry, and may be the cause of the intermediate periodicity detected in some cases.

A slightly different situation can arise if there is even a slight misalignment between the disk and the binary orbit. In these cases, tidally induced disk warping can occur, and will cause disk precession \citep{Martin2011}. Spectroscopic evidence for a warped disk can be seen in Be star systems that transition between a shell-absorption phase and an emission phase. 

\citet{Sterken1996} investigate intermediate periodicity in a sample of 15 Be stars with spectral types earlier than B3, finding four (27$\%$) with periods between 4 - 93 days. The authors suggest pulsations, rotation of an inhomogeneous stellar surface, and/or oscillations in the circumstellar envelope as plausible explanations for the shorter period case (HD 89890, P=4.656 days). For the three with longer periods (HD 173219, P=61.4 days; HD 48917, P=87.9 days; HD 58978, P=92.7 days), an elliptical precessing disk is suggested. One of these, HD 173219, has a confirmed radial velocity orbit within $\sim$5\% of the photometric period \citep{Hutchings1973}, which supports the idea that tidal forces acting on the circumstellar disk as a result of a binary companion can be responsible for the observed periodicity.

Another scenario involving brightness modulated by binarity is the well known case of ellipsoidal variables, where one or both binary components are elongated according to the gravitational influence of the other component. The star HD 50123 (misclassified as a classical Be star in BeSS) was found by \citet{Sterken1994} to be an interacting binary consisting of a B6Ve primary and an early K giant secondary filling its Roche lobe, with each of the two components contributing roughly the same to the total flux in the V-band. This binary system is at an intermediate inclination angle (on the order of 60$^\circ$), has a mass ratio of q$\approx$0.3, an orbital period of 28.601 days, and shows a double-waved modulation of its light curve typical of ellipsoidal variables. In this configuration, gas from the K giant feeds the circumstellar shell around the primary component. \citet{Rivinius2013} states that no classical Be stars with a Roche lobe filling companion are known, since mass transferring binaries are excluded from the definition of classical Be stars. Although HD 50123 is not a classical Be star, the accretion onto the B-type star results in similar observable features, namely H$\alpha$ emission and rapid rotation. It is therefore likely that some of the IP variables presented here are not classical Be stars, despite the possible presence of line emission and a B-type spectral designation. A more detailed analysis would be required to make any claims regarding the classical Be star status of these objects. 

\section{Conclusion}
\label{sec:conclusion}

\begin{table*}
 \centering
 \caption{Fractions showing variability}
 \label{tbl:variable_fractions}
 \begin{tabular}{|l|l|l|l|l|l|}
    \hline
    \hline
 \textit{Variable Type} & \textit{Description of Variability} & \textit{All} & \textit{Early} & \textit{Mid + Late} & \textit{Unclassified}\\

\hline

Outburst Variation (ObV) & One or more outburst & 36$\%$ (\(\nicefrac{168}{470}\))& 51$\%$ (\(\nicefrac{135}{265}\)) & 12$\%$ (\(\nicefrac{16}{139}\)) & 26$\%$ (\(\nicefrac{17}{66}\)) \\
\hline

Semi-Regular Outbursts (SRO) & Outbursts occurring at regular intervals & 21$\%$ (\(\nicefrac{35}{168}\)) & 22$\%$ (\(\nicefrac{29}{135}\)) & 19$\%$ (\(\nicefrac{3}{16}\)) & 18$\%$ (\(\nicefrac{3}{17}\)) \\
\hline

Long-Term Variation (LTV) & Variability on timescales of years & 37$\%$ (\(\nicefrac{81}{217}\)) & 45$\%$ (\(\nicefrac{54}{121}\)) & 21$\%$ (\(\nicefrac{15}{73}\)) & 52$\%$ (\(\nicefrac{12}{23}\))\\
\hline

Non-Radial Pulsations (NRP) & Periodic variation, $P$ $\leq$ 2 days & 25$\%$ (\(\nicefrac{125}{510}\))& 28$\%$ (\(\nicefrac{80}{287}\))& 21$\%$ (\(\nicefrac{32}{155}\)) & 19$\%$ (\(\nicefrac{13}{68}\))  \\
\hline

Intermediate Periodicity (IP) & Periodic variation, 2 $<$ $P$ $\leq$ 200 days & 39$\%$ (\(\nicefrac{201}{510}\)) & 52$\%$ (\(\nicefrac{150}{287}\)) & 21$\%$ (\(\nicefrac{32}{155}\)) & 28$\%$ (\(\nicefrac{19}{68}\))  \\
\hline

Eclipsing Binaries (EB) & Eclipsing systems & 2.7$\%$ (\(\nicefrac{14}{510}\)) & 2.8$\%$ (\(\nicefrac{8}{287}\)) & 1.9$\%$ (\(\nicefrac{3}{155}\)) & 4.4$\%$ (\(\nicefrac{3}{68}\)) \\

   \hline
    \hline
 \end{tabular}
 \begin{flushleft}
  \footnotesize 
  \footnotesize Fraction of stars showing each type of variability, according to their spectral type. The category `all' includes early-, mid-, and late-type stars, as well as those unclassified in BeSS. The fraction of systems with SROs is calculated from the subset of stars showing at least one outburst, and EBs from all non-saturated objects. See Section~\ref{sec:results} for an explanation of how the other fractions were calculated. 
  \end{flushleft}  
\end{table*}  

\vspace{2mm}

\begin{figure}[!ht]
\centering\epsfig{file=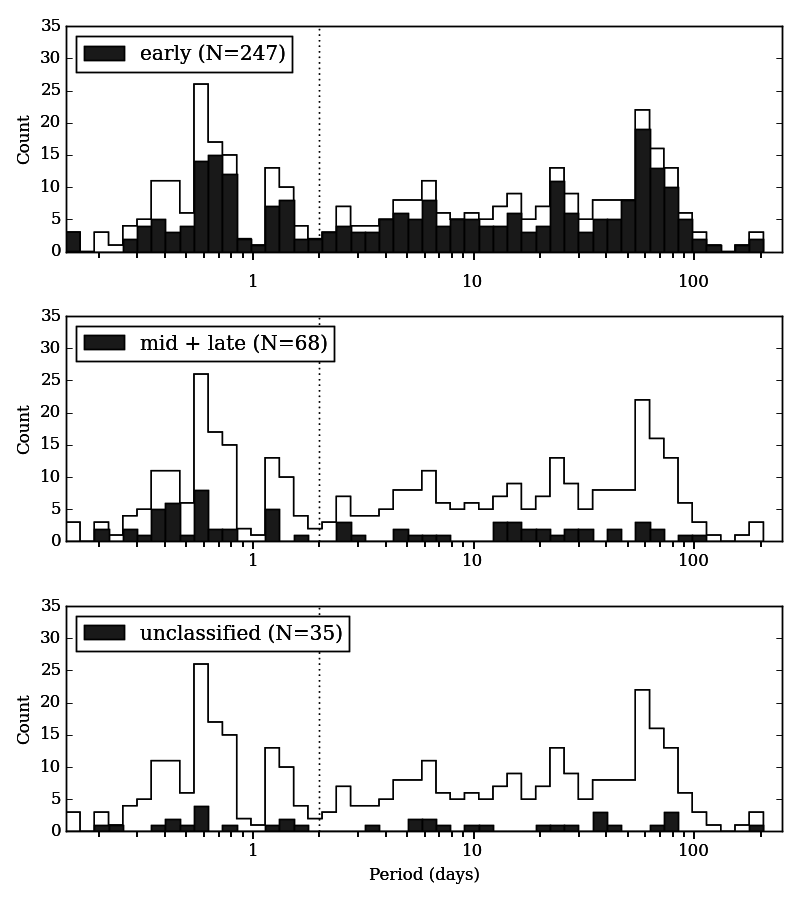,clip=,width=0.99\linewidth}
\caption{Histogram showing the distribution of all periodic signals found in objects within the BeSS-KELT sample on a logarithmic scale spanning between 0.1 - 200 days, displayed in the same manner as in Figure~\ref{fig:Outburst_rates_early}. The vertical dotted line at 2 days marks the cutoff between short periods and intermediate periods. The short-period variables are left of the dotted line, and are best explained by NRP modes in the star. All types of periodic variability (NRP, IP, SRO, and EB) are included in this histogram. A single star may have multiple periods at different timescales and thus may appear in up to three different bins, although the majority of periodically variable stars have a single dominant frequency.}
\label{fig:hist_all_periodic}
\end{figure}

The fractions of stars in this sample showing each type of variability mentioned in Table~\ref{tbl:variable_class} are summarized in Table~\ref{tbl:variable_fractions}. There are many cases where a star exhibits both non-periodic (e.g. outbursts) and periodic variability (e.g. pulsations), and it is important to note that these categories are not mutually exclusive. 

A histogram showing the distribution of all recovered periods is shown in Figure~\ref{fig:hist_all_periodic}. This histogram includes all types of periodic variability (NRP, IP, SRO, and EB variables), and is complicated by the fact that a single star can exhibit more than one period in its light curve. Because of the diurnal sampling of the KELT survey, periodic signals very close to one day are poorly sampled. The dearth of detected periods very near to one day is a result of this systematic effect. 

From analyzing the KELT light curves of this sample of Be stars, we arrive at a few important conclusions. Consistent with other studies \citep[e.g.][]{Cuypers1989,Gutierrez-Soto2008,McSwain2009,Chojnowski2015}, we find that Be stars are a highly variable class of objects, with a greater degree of variability seen in earlier spectral types. About 1/5 of Be stars with clear photometric outbursts have them occurring at semi-regular intervals. Intermediate periodicity (longward of 2 days) is a common occurrence, and is seen in 41$\%$ of our sample. By combining KELT data with BeSS spectra, we provide evidence that photometric outbursts correspond to disk creation or disk building events, and that global disk oscillations manifesting in V/R variability can also modulate the brightness of a Be star + disk system.

This work is unique in its large sample size of Galactic Be stars and long temporal coverage, both of which will continue to grow as new KELT data is collected and reduced. Future work will involve increasing our statistics with larger sample sizes and baselines, as well as more detailed investigations of particularly interesting systems and a more thorough treatment of the types of variability discussed here.

\section{Acknowledgements}
This work has made use of the BeSS database, operated at LESIA, Observatoire de Meudon, France: http://basebe.obspm.fr. M.V.M. acknowledges support from NSF grant AST - 1109247. Work by D.J.S and B.S.G was partially supported by NSF CAREER Grant AST-1056524. J.E.B. and K.S.B. acknowledge support from NSF grant AST-1412135. This research has made use of the SIMBAD database, operated at CDS, Strasbourg, France. J.L.-B. acknowledges institutional support from Lehigh University, and a Sigma Xi Grant-in-Aid of research. This research has made use of NASA's Astrophysics Data System. 

\appendix

\begin{figure}[!ht]
\centering\epsfig{file=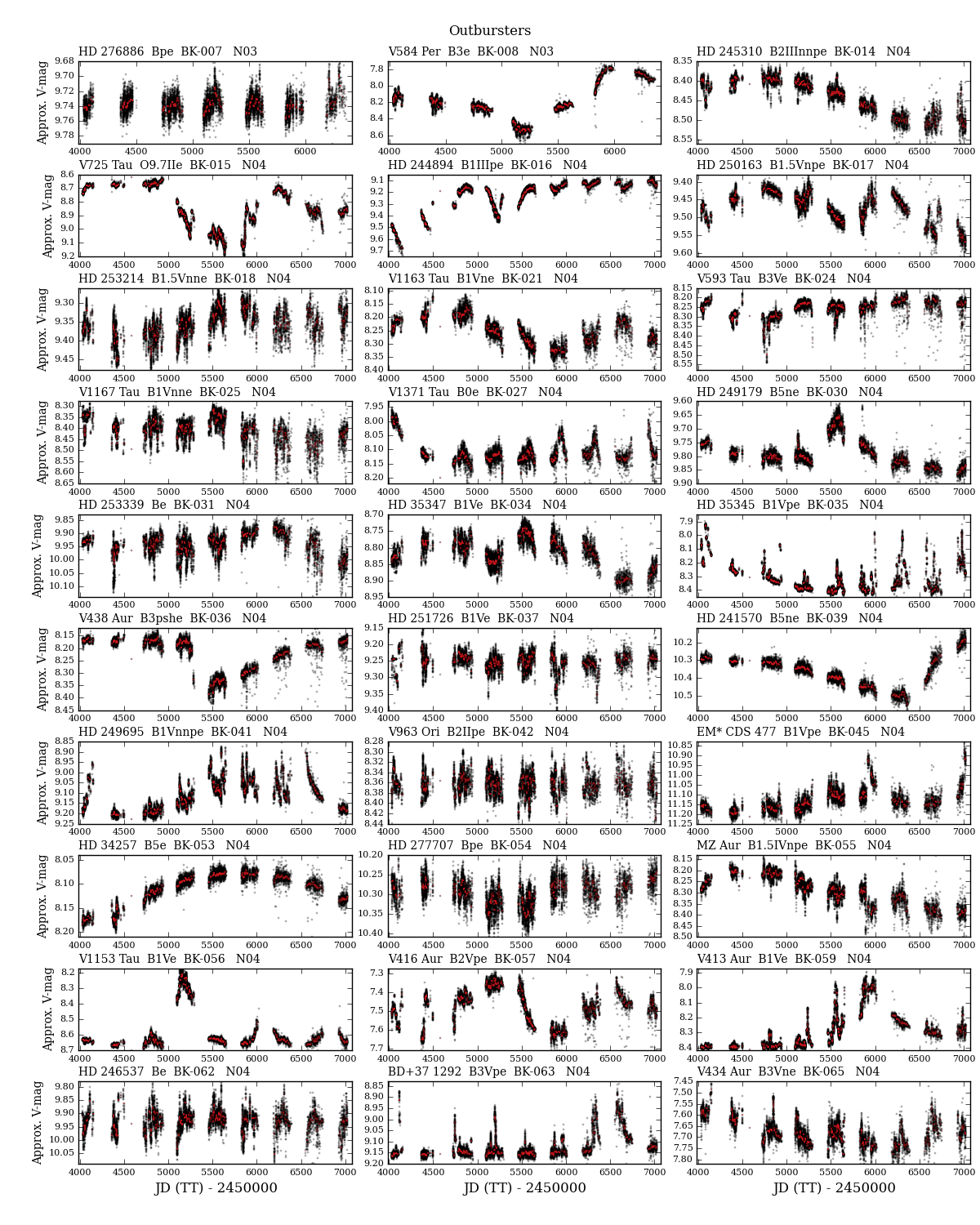,clip=,width=0.99\linewidth}
\caption{Raw KELT light curves for all objects exhibiting outbursts. Black points are raw data, and red points show a version of the light curve with a low-pass filter applied. Subtitles read: object identifier, literature spectral type, BeSS-KELT number, and KELT field. }
\label{fig:Composite_ObV_0}
\end{figure}

\begin{figure}[!ht]
\centering\epsfig{file=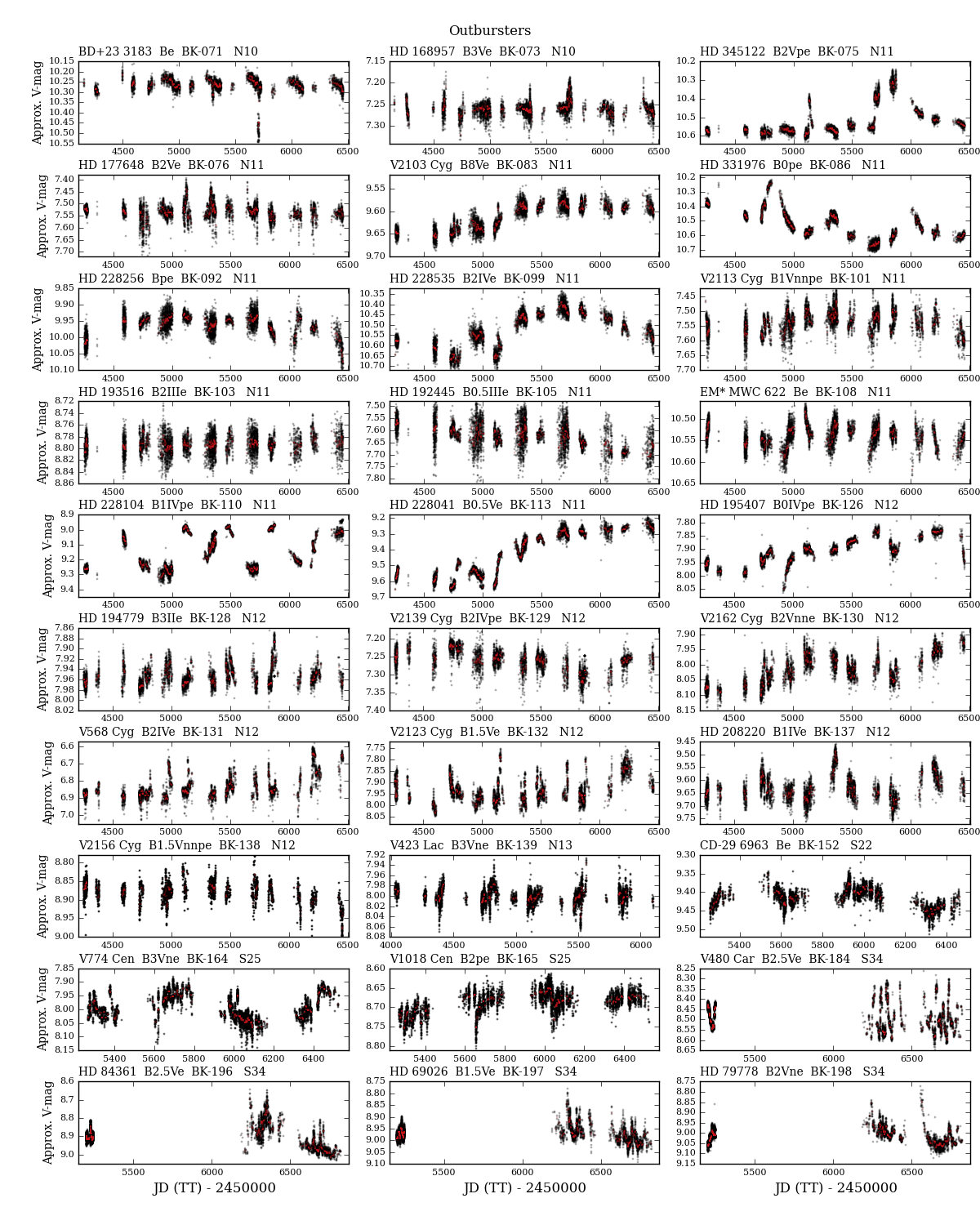,clip=,width=0.99\linewidth}
\caption{Same as Figure~\ref{fig:Composite_ObV_0}}
\label{fig:Composite_ObV_1}
\end{figure}

\begin{figure}[!ht]
\centering\epsfig{file=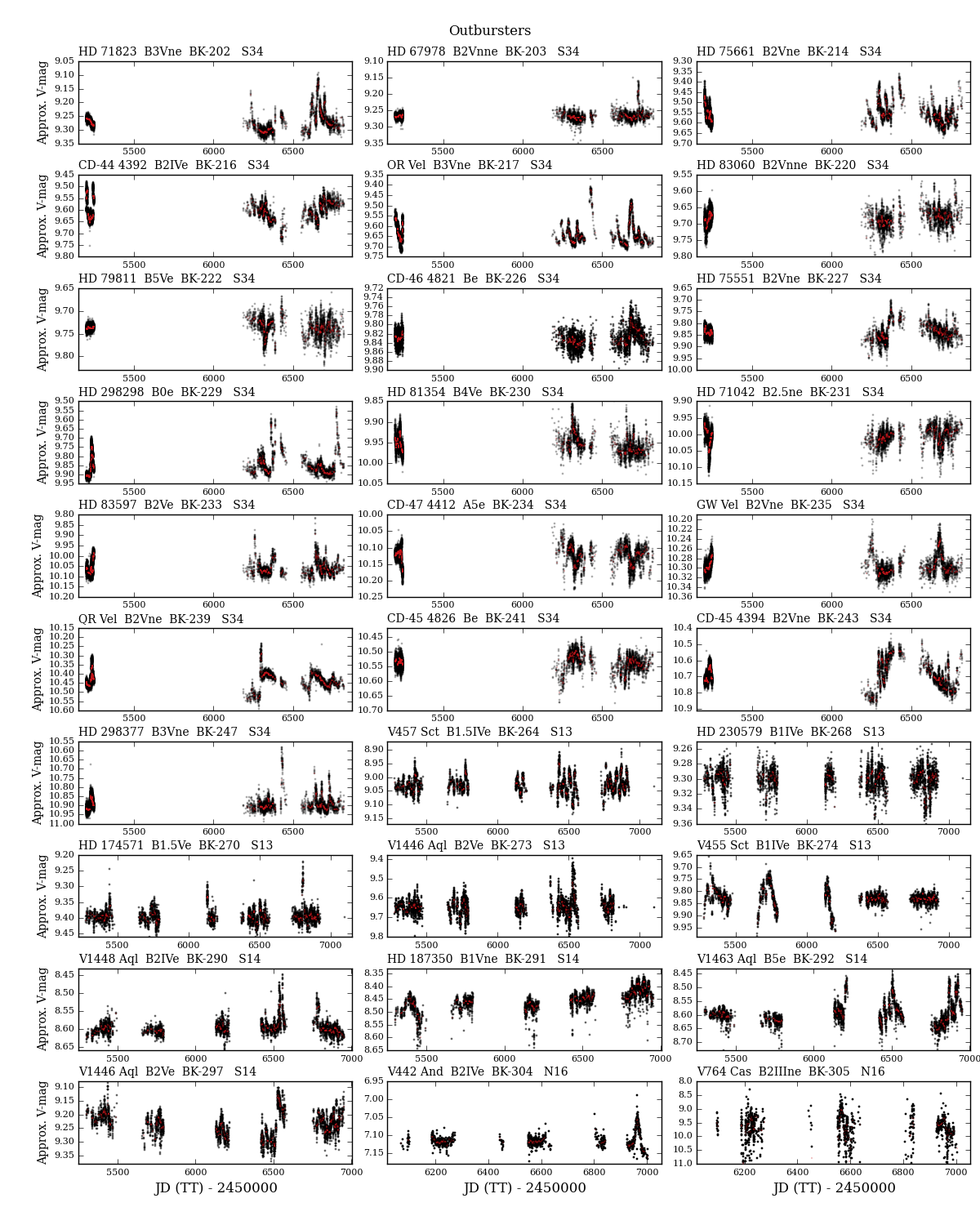,clip=,width=0.99\linewidth}
\caption{Same as Figure~\ref{fig:Composite_ObV_0}}
\label{fig:Composite_ObV_2}
\end{figure}

\begin{figure}[!ht]
\centering\epsfig{file=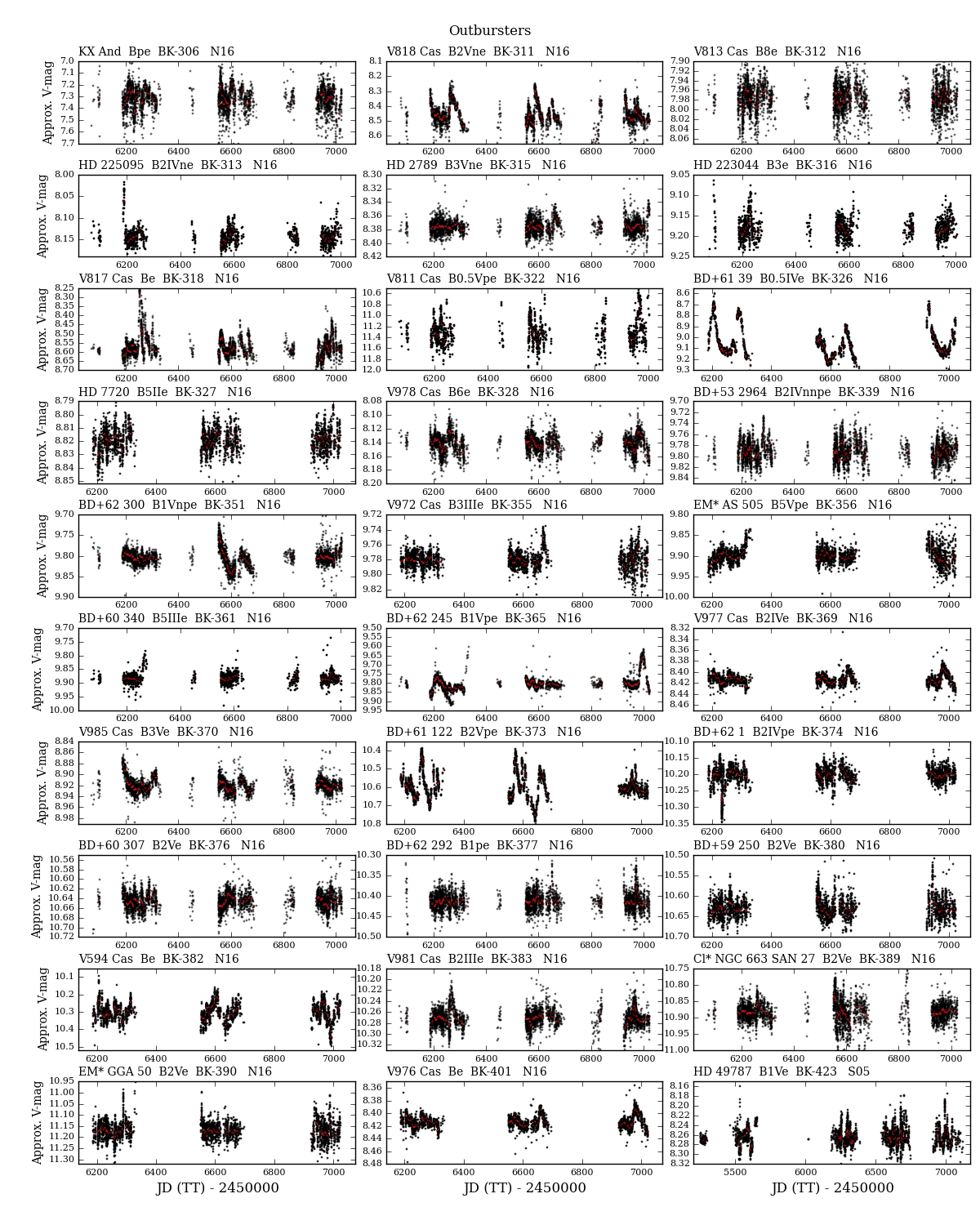,clip=,width=0.99\linewidth}
\caption{Same as Figure~\ref{fig:Composite_ObV_0}}
\label{fig:Composite_ObV_3}
\end{figure}

\begin{figure}[!ht]
\centering\epsfig{file=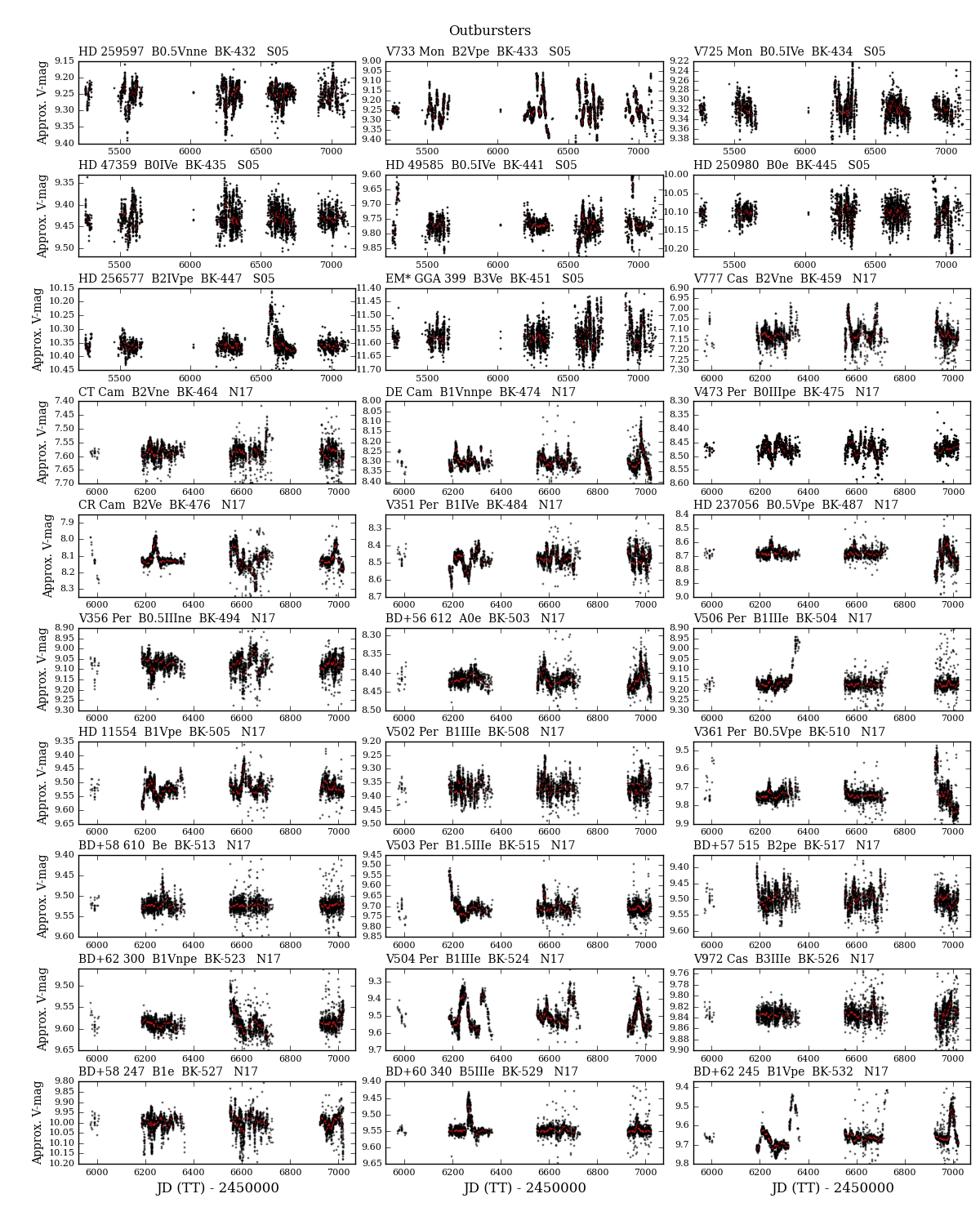,clip=,width=0.99\linewidth}
\caption{Same as Figure~\ref{fig:Composite_ObV_0}}
\label{fig:Composite_ObV_4}
\end{figure}

\begin{figure}[!ht]
\centering\epsfig{file=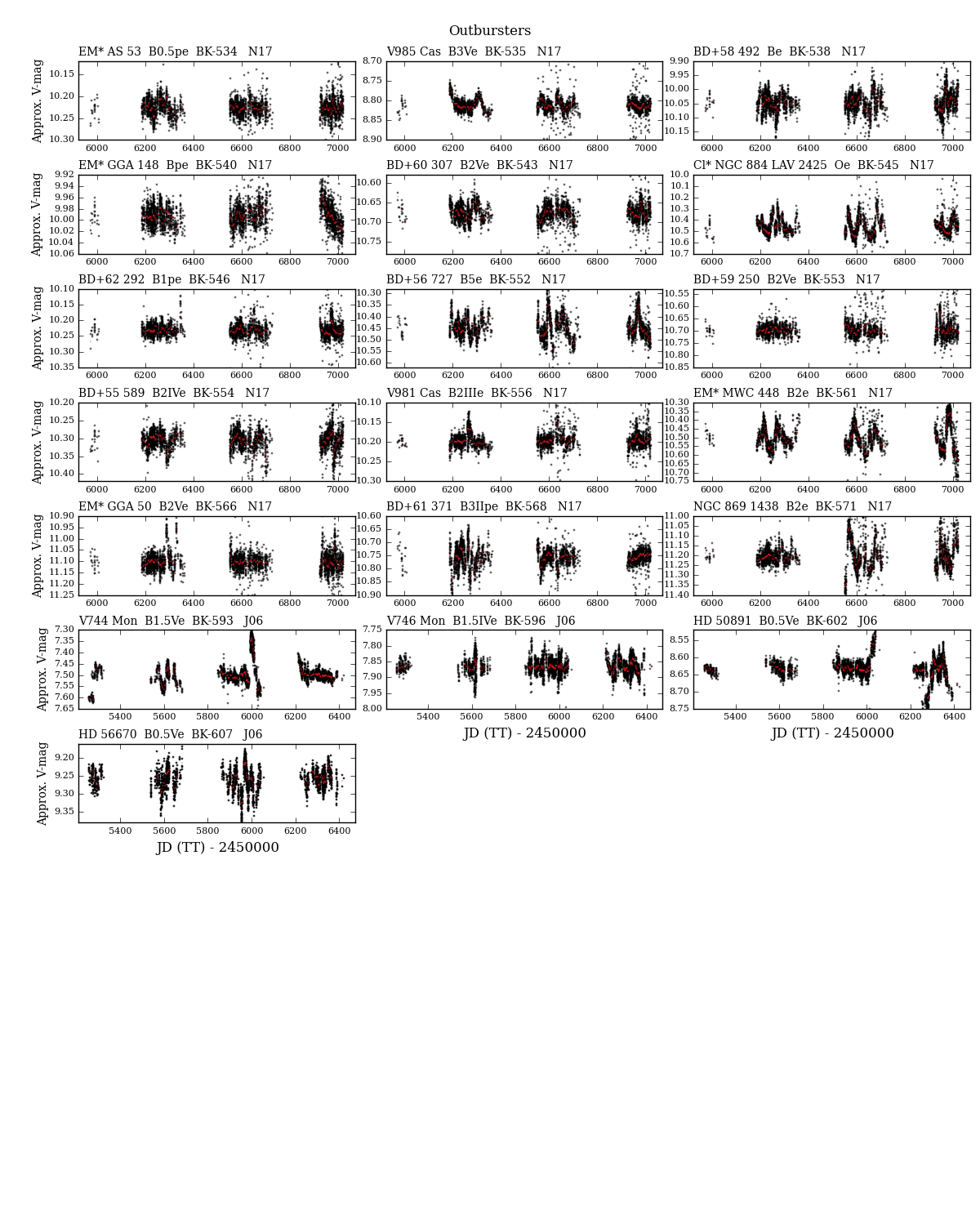,clip=,width=0.99\linewidth}
\caption{Same as Figure~\ref{fig:Composite_ObV_0}}
\label{fig:Composite_ObV_5}
\end{figure}

\begin{figure}[!ht]
\centering\epsfig{file=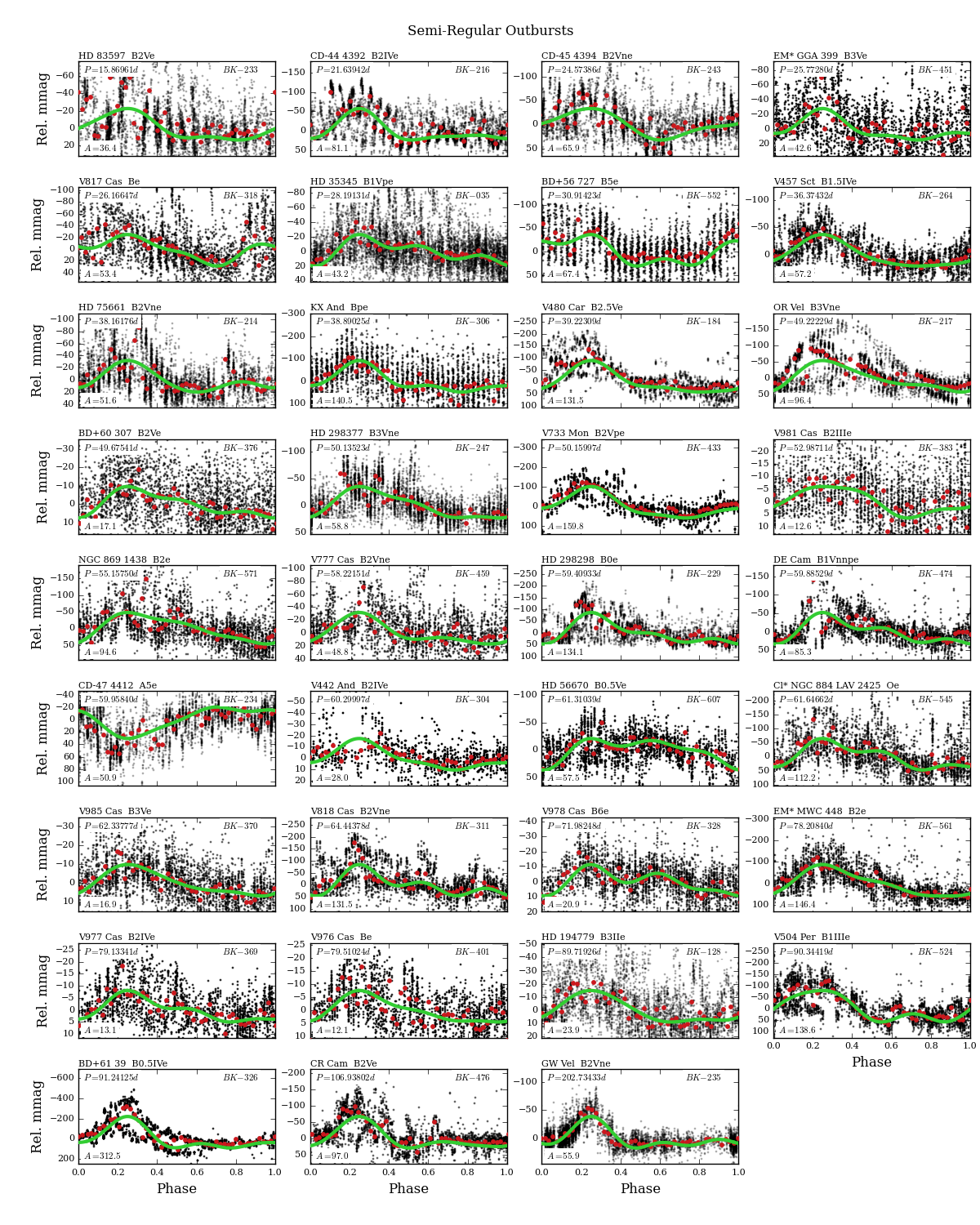,clip=,width=0.99\linewidth}
\caption{Phased KELT light curves for all objects exhibiting SRO, displayed in order of increasing period. Red points show the data median-binned with a bin size of 0.02 in phase. The solid green line shows a fit to the data using a combination of three sinusoids, to guide the eye. In each sub-plot, The object identifier and spectral type are printed in the title, the period in the upper left corner, the BeSS-KELT number in the upper right corner, and the $max-min$ amplitude of the fit, in mmag, is shown in the bottom left corner.}
\label{fig:Composite_SRO_0}
\end{figure}

\begin{figure}[!ht]
\centering\epsfig{file=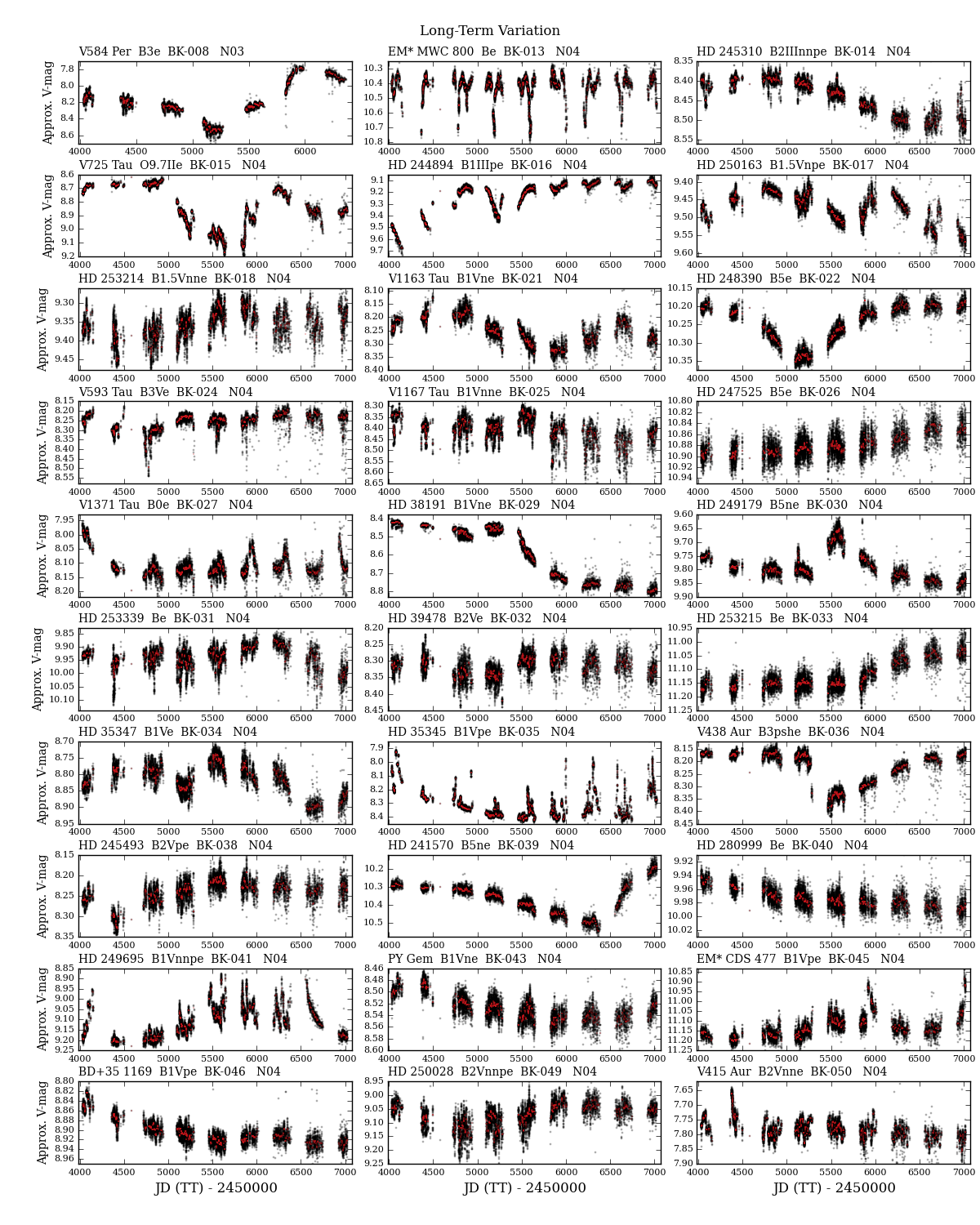,clip=,width=0.99\linewidth}
\caption{Same as Figure~\ref{fig:Composite_ObV_0}, but for LTV.}
\label{fig:Composite_LTV_0}
\end{figure}

\begin{figure}[!ht]
\centering\epsfig{file=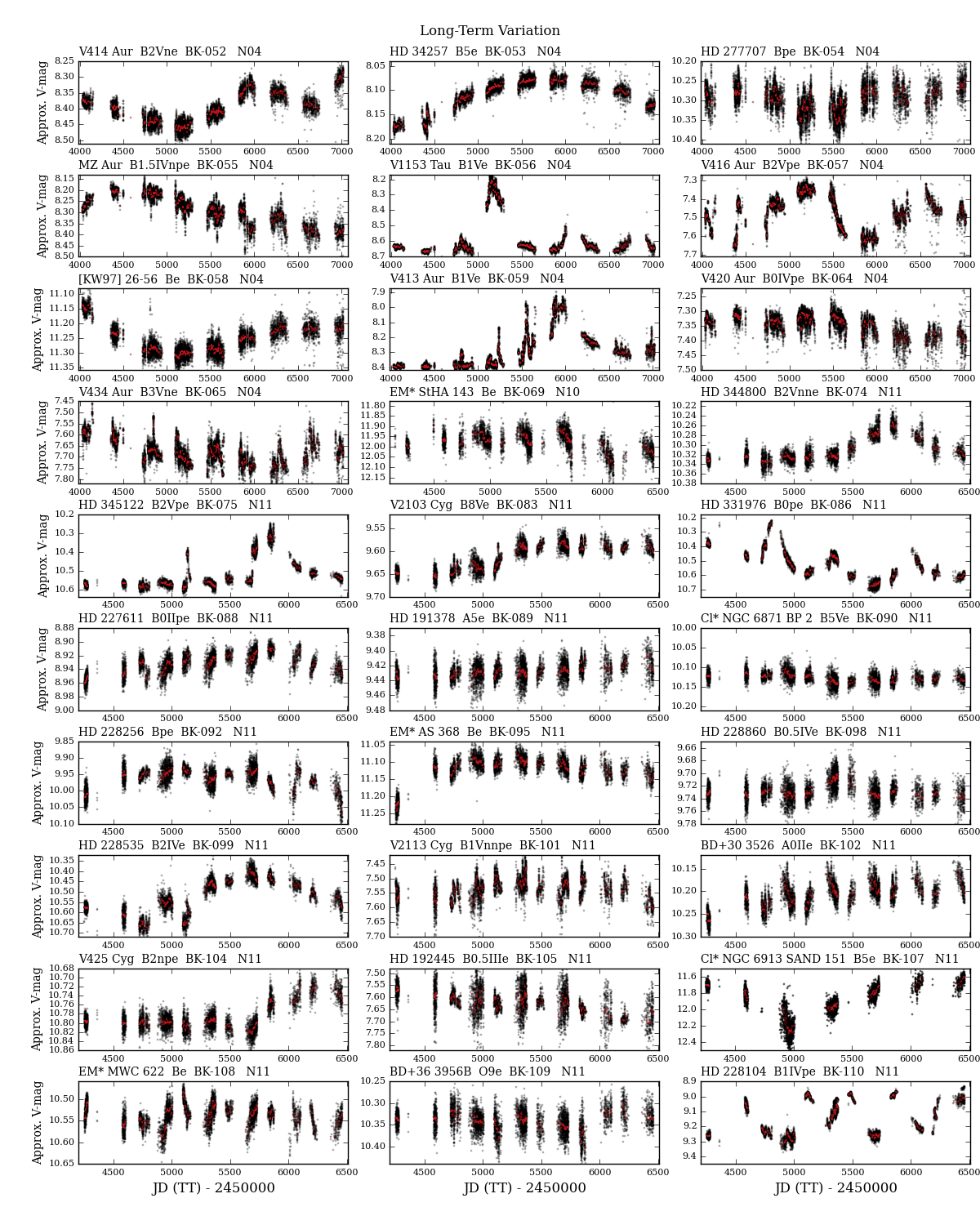,clip=,width=0.99\linewidth}
\caption{Same as Figure~\ref{fig:Composite_LTV_0}}
\label{fig:Composite_LTV_1}
\end{figure}

\begin{figure}[!ht]
\centering\epsfig{file=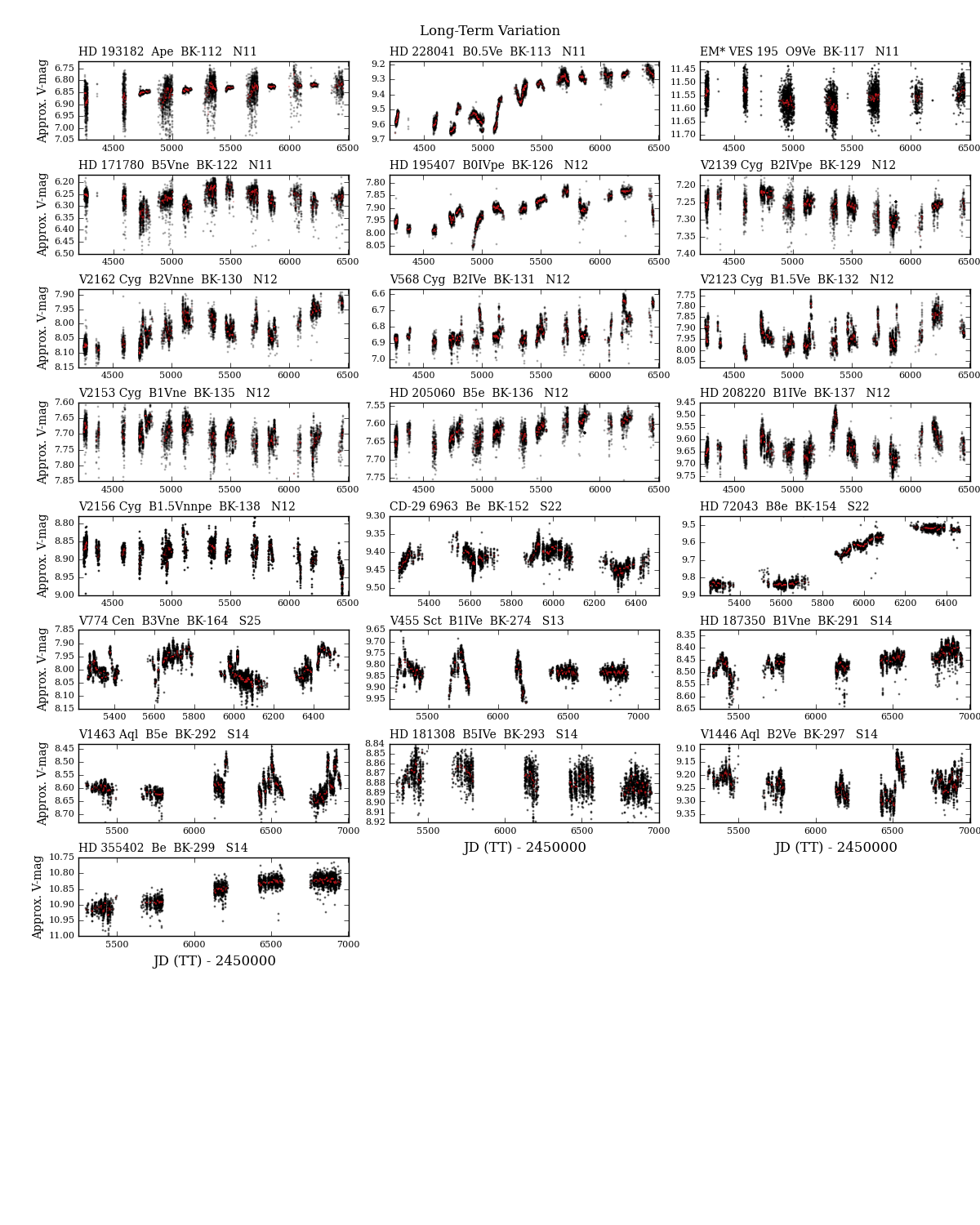,clip=,width=0.99\linewidth}
\caption{Same as Figure~\ref{fig:Composite_LTV_0}}
\label{fig:Composite_LTV_2}
\end{figure}

\begin{figure}[!ht]
\centering\epsfig{file=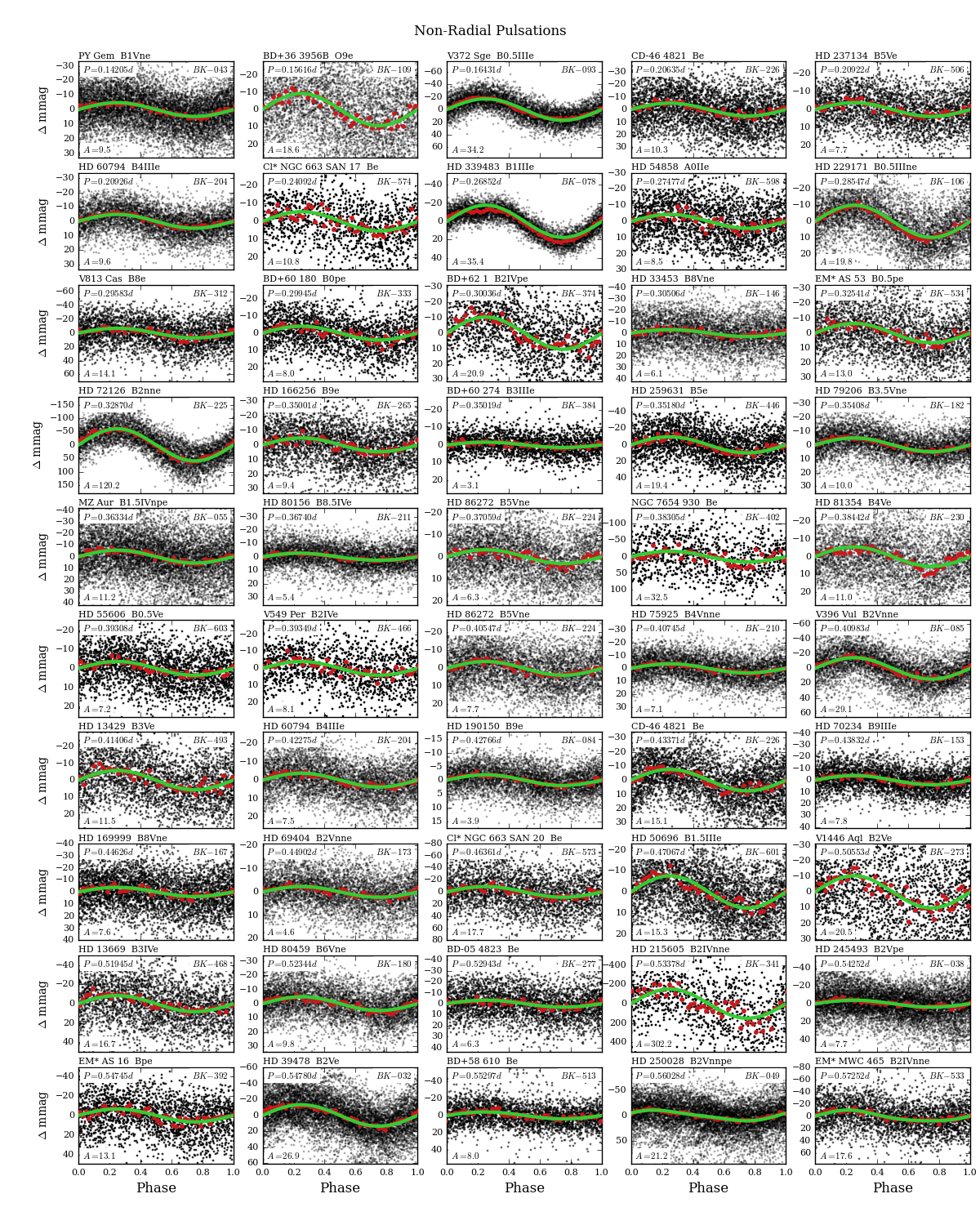,clip=,width=0.99\linewidth}
\caption{Same as Figure~\ref{fig:Composite_SRO_0}, but for NRP.}
\label{fig:Composite_NRP_0}
\end{figure}

\begin{figure}[!ht]
\centering\epsfig{file=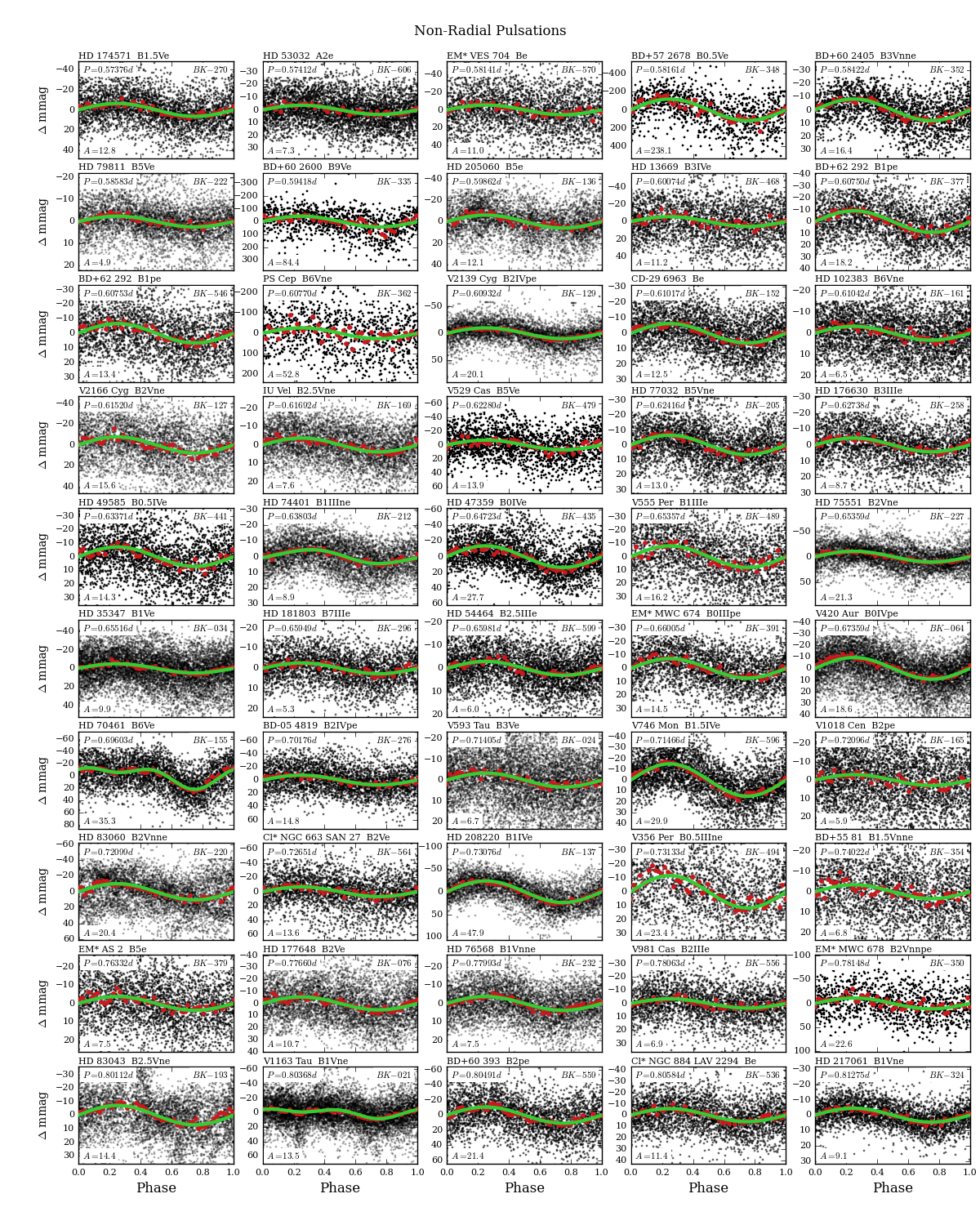,clip=,width=0.99\linewidth}
\caption{Same as Figure~\ref{fig:Composite_NRP_0}.}
\label{fig:Composite_NRP_1}
\end{figure}

\begin{figure}[!ht]
\centering\epsfig{file=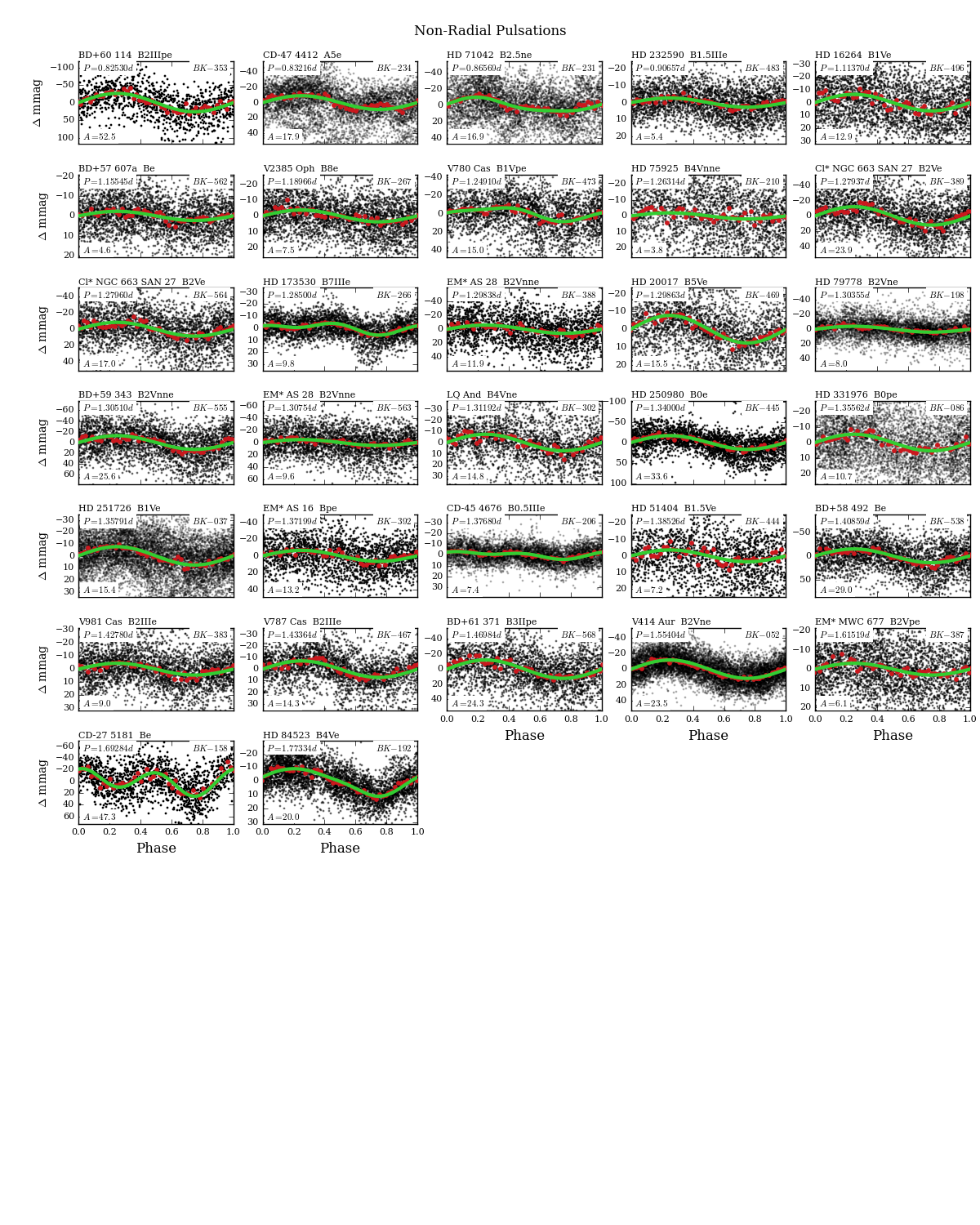,clip=,width=0.99\linewidth}
\caption{Same as Figure~\ref{fig:Composite_NRP_0}.}
\label{fig:Composite_NRP_2}
\end{figure}

\begin{figure}[!ht]
\centering\epsfig{file=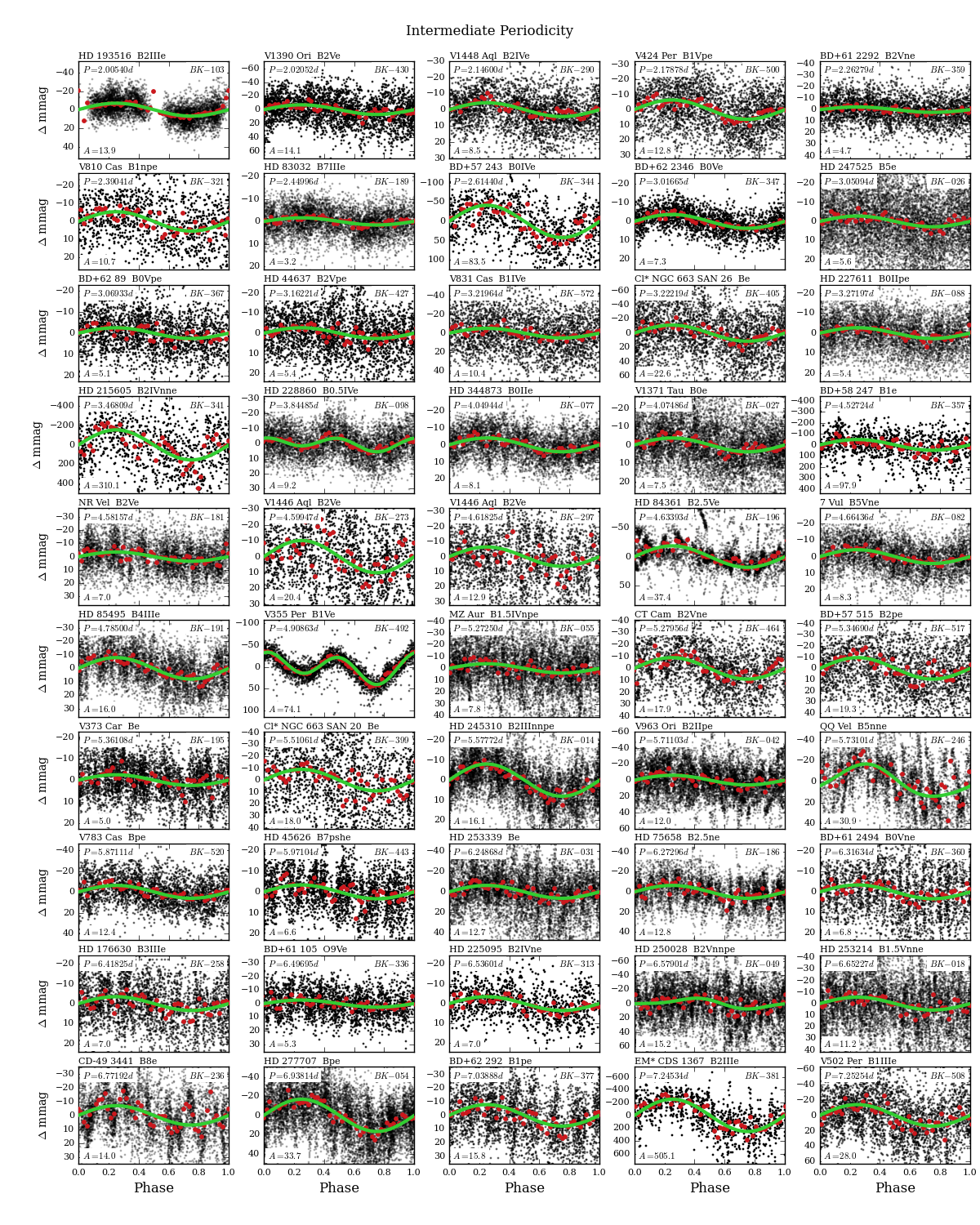,clip=,width=0.99\linewidth}
\caption{Same as Figure~\ref{fig:Composite_SRO_0}, but for IP (excluding EB or SRO) variables.}
\label{fig:Composite_IP_0}
\end{figure}

\begin{figure}[!ht]
\centering\epsfig{file=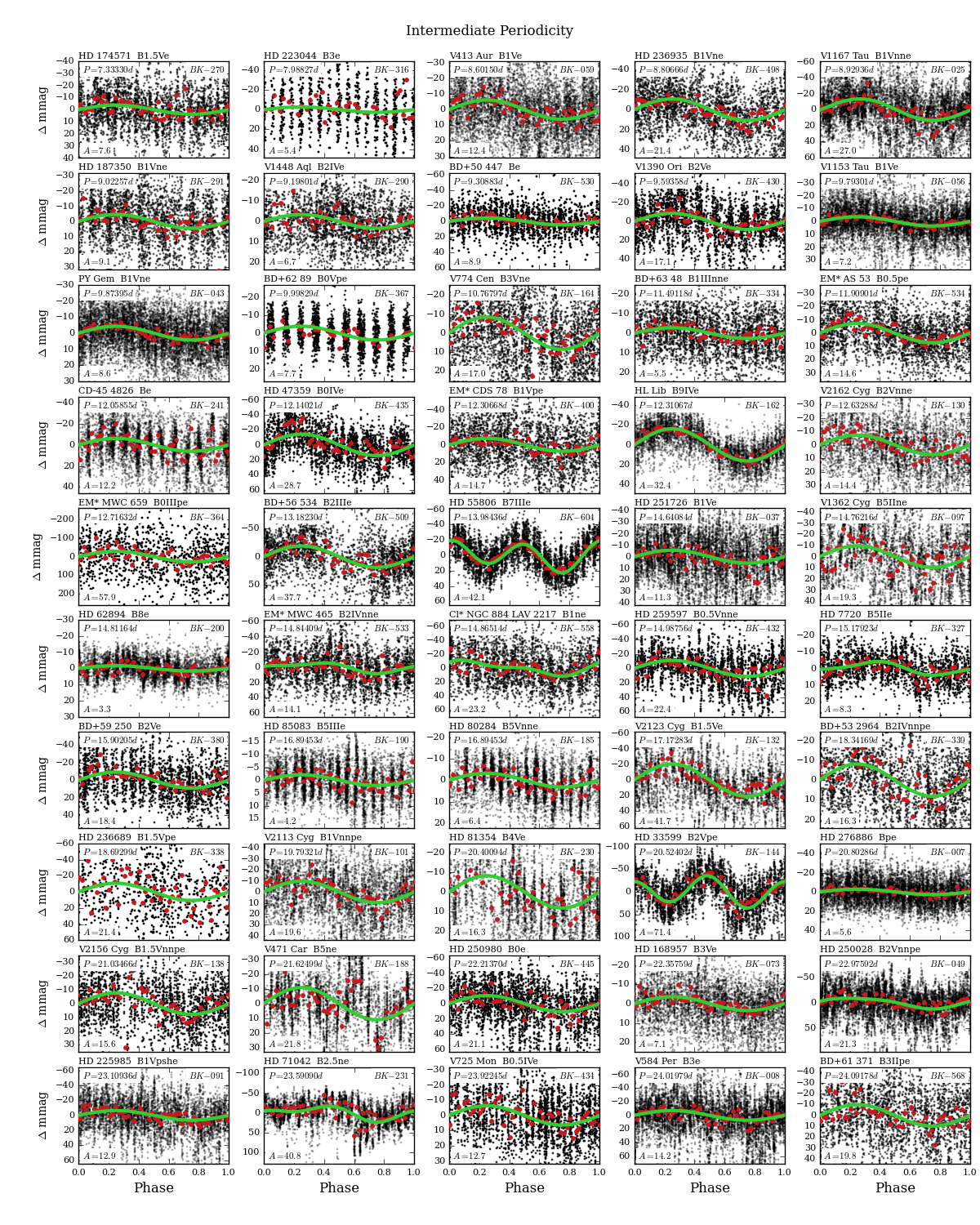,clip=,width=0.99\linewidth}
\caption{Same as Figure~\ref{fig:Composite_IP_0}.}
\label{fig:Composite_IP_1}
\end{figure}

\begin{figure}[!ht]
\centering\epsfig{file=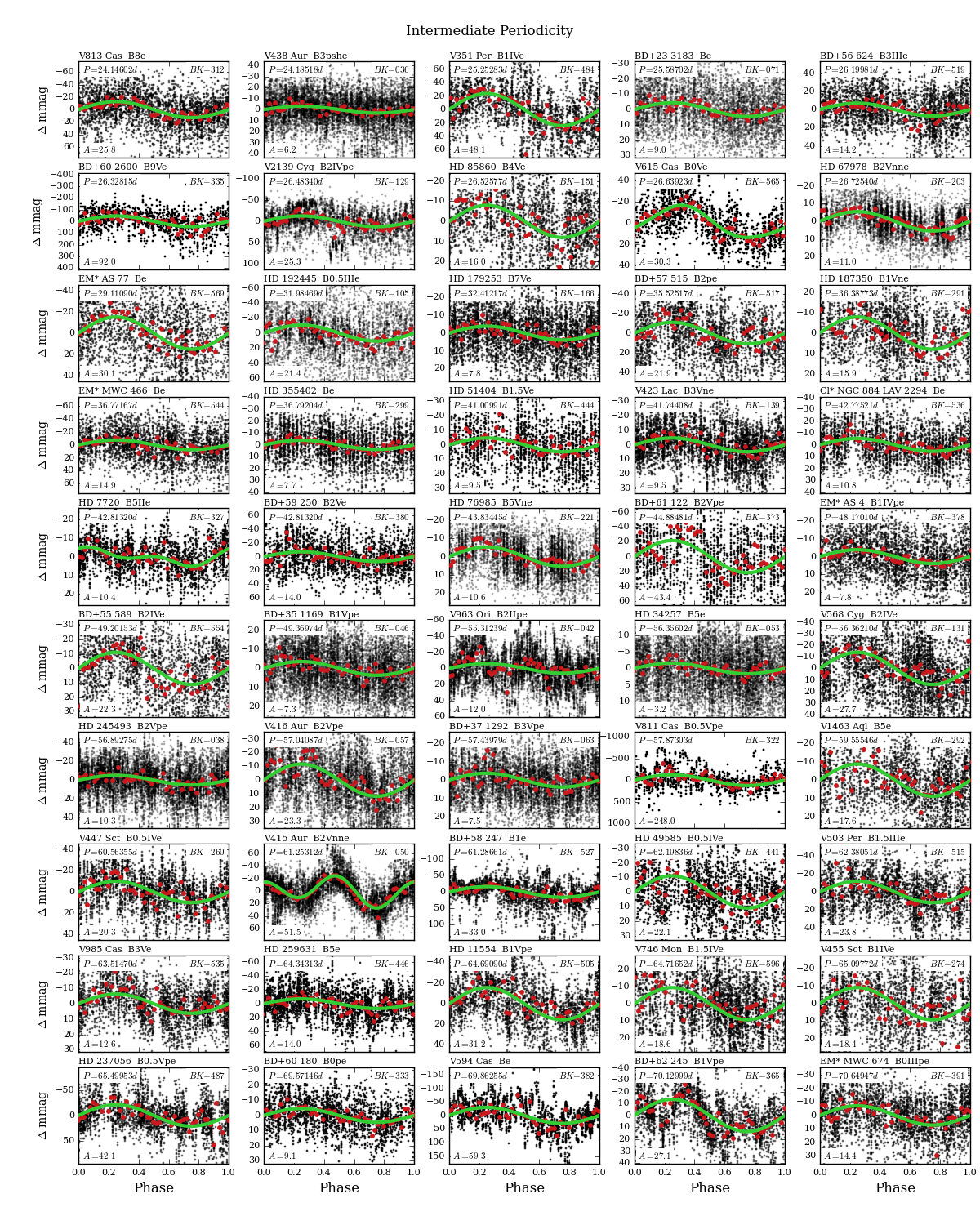,clip=,width=0.99\linewidth}
\caption{Same as Figure~\ref{fig:Composite_IP_0}.}
\label{fig:Composite_IP_2}
\end{figure}

\begin{figure}[!ht]
\centering\epsfig{file=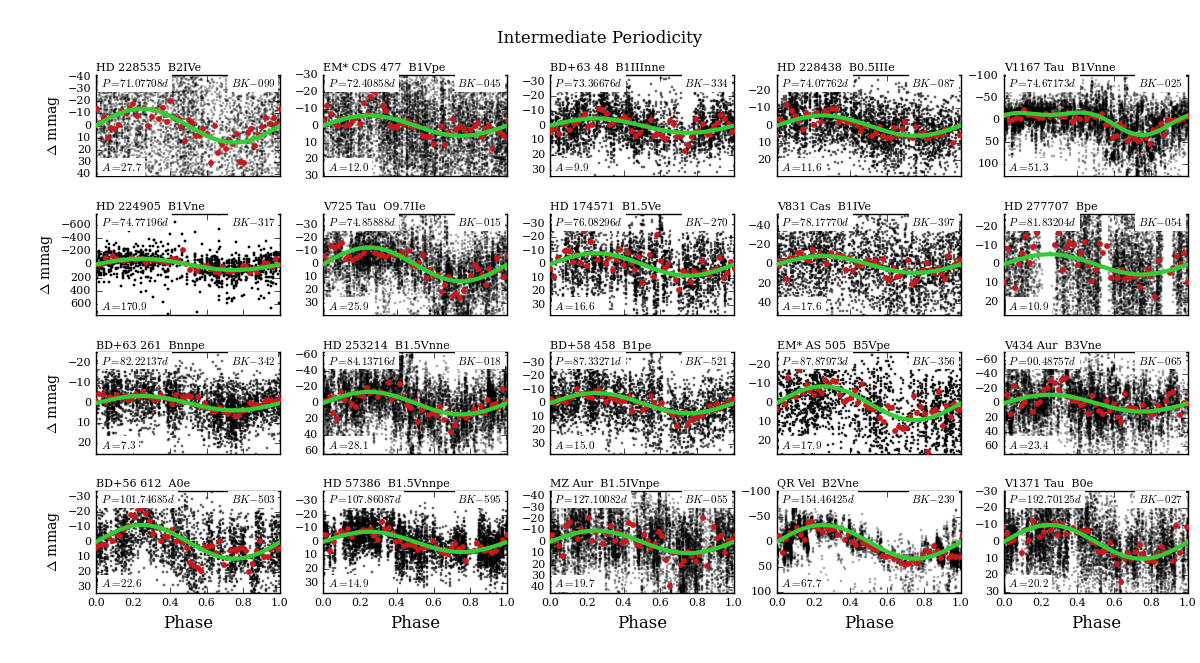,clip=,width=0.99\linewidth}
\caption{Same as Figure~\ref{fig:Composite_IP_0}.}
\label{fig:Composite_IP_3}
\end{figure}

\newpage

\section{Eclipsing Binaries}
\begin{figure}[!ht]
\centering\epsfig{file=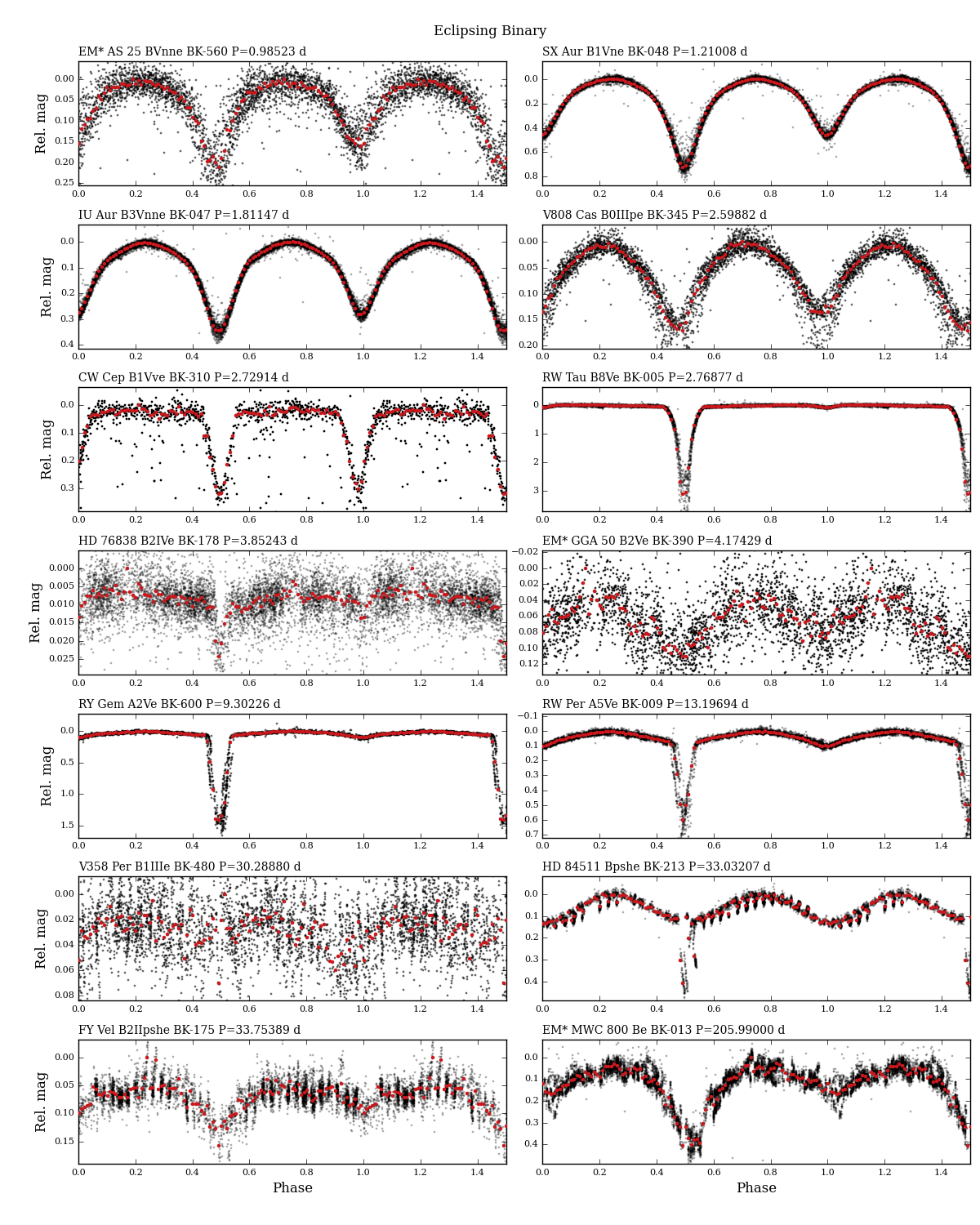,clip=,width=0.99\linewidth}
\caption{Phased KELT light curves for all objects identified as EBs, displayed in order of increasing period. Red points show the data median-binned with a bin size of 0.01 in phase. The object identifier, spectral type, BeSS-KELT number, and period are printed above each sub-plot.}
\label{fig:Composite_EB_0}
\end{figure}

We identify 14 stars from the BeSS-KELT sample as eclipsing binaries (EBs). The orbital periods of these systems range from 0.98523 days to 206 days, and show a wide range of morphologies. Some of these systems have been studied extensively in the literature (e.g. BK-048 = HD 33357), while others are newly discovered EBs. Brief descriptions of these systems are given below, ordered by increasing orbital period. Plots of phased light curves for these systems are displayed in Figure~\ref{fig:Composite_EB_0}. \vspace{2mm} \\ 
\textit{BK-560 = EM* AS 25 ($P=0.98523$ $d$)}\\
This is not listed in the literature as an EB, but exists in catalogs of stars with H$\alpha$ emission \citep{Kohoutek1999}. No BeSS spectra are available. Given the short orbital period derived from KELT photometry (P=0.98523 days), this is not a classical Be star with a decretion disk, but is more likely to be a binary system where one of the two components is accreting mass from the other. It is this accretion disk giving rise to the Hydrogen emission.  \vspace{2mm} \\ 
\textit{BK-048 = HD 33357 ($P=1.21008$ $d$):}\\
This star is listed in BeSS as having spectral type B1Vne, although none of the contributed BeSS spectra for this object show H$\alpha$ in emission. This has been known to be an EB since at least 1938 \citep{Luyten1938}. This system was recently studied by \citet{Ozturk2014}, who modeled the orbital and stellar parameters of the system. They conclude that the system is a rare example of a binary pair at the border between semi-detached and contact phases with the primary and secondary components having spectral types of B2V and B4V respectively. Furthermore, they determine that the orbital period is increasing at a rate of 0.0055 $s/yr$ as a result of non-conservative mass transfer from the secondary to the primary component. \vspace{2mm} \\ 
\textit{BK-047 = V* IU Aur = HD 35652 ($P=1.81147$ $d$):}\\
This star is listed in BeSS as having a spectral type of B3Vnne, although none of the contributed BeSS spectra for this object show H$\alpha$ in emission. This system is a known $\beta$ Lyra type EB. \citet{Ozdemir2003} analyze the timing of this system and find evidence for a third component orbiting the system on an eccentric orbit ($e$ = 0.62) with a period of 293.3 days, noting that the third body is likely a binary pair itself. \vspace{2mm} \\ 
\textit{BK-463 = V782 Cas = HD 12882 ($P=2.51317$ $d$):}\\
Many BeSS spectra for this object clearly show H$\alpha$ in emission. The morphology of the H$\alpha$ line is unusual, showing a single peaked line that is flat near the peak intensity, rather than being sharply pointed. A literature search indicates that this is not a previously known EB, although \citet{Lefevre2009} list this as a possible variable star (``ACYG?'' - this is attributed to variable stars of spectral class Ia-Iab-Ib, I/II), with a period of 47.621 days, an amplitude of 0.081 mag and a spectral type of B6Ia. \vspace{2mm} \\
\textit{BK-345 = BD+61 2408 = V808 Cas ($P=2.59882$ $d$):}\\
This star is listed in BeSS as having a spectral type B0IIIpe, in Simbad as a B3 star, and is classified as a B0.2IV star by \citet{Negueruela2004}. The two contributed BeSS spectra do not show emission features, and it is unclear where this has been identified as an emission line star. \citet{Lefevre2009} list this as a variable OB star based on Hipparcos photometry. More specifically, they classify V808 Cas as a SPB type star with a period of 1.300 days and an amplitude of 0.187 mag. This is half the period we find using KELT data, but we classify this system as an EB based on the morphology of the phased light curve, which shows two distinct eclipses with different depths. A literature search gives no indication of V808 Cas being a classical Be star, nor is this system known to be an EB.  \vspace{2mm} \\ 
\textit{BK-310 = CW Cep = HD 218066 ($P=2.72914$ $d$):}\\
This object is a known early-type double-lined EB with apsidial motion indicating a third companion \citep[e.g.][]{Wolf2006}. Many BeSS spectra show clear emission, although it is unclear if the emission can be attributed to a decretion disk typical of a classical Be star. \vspace{2mm} \\ 
\textit{BK-005 = RW Tau = HD 25487 ($P=2.76877$ $d$):}\\
Although listed as a Be star in BeSS, this system is not a classical Be star. \citet{Vesper1993} present evidence that this is a B8V + K0IV system, noting six different explanations for the H$\alpha$ emission (their section 7), none of which is a `normal' Be star decretion disk. The H$\alpha$ line is modulated by the orbital period, and the most likely explanation is the presence of an accretion disk where mass is being transferred to the B-type component. The primary eclipse dips below KELT's faintness threshold, but appears to be at least three magnitudes deep in KELT's bandpass. \vspace{2mm} \\
\textit{BK-178 = HD 76838 ($P=3.85243$ $d$):}\\
Listed in BeSS as having a spectral type of B2IVe, this star is in the Jaschek catalog of Be stars \citep{Jaschek1982}. No BeSS spectra are available for this star. A literature search indicates this is not a known EB. \vspace{2mm} \\ 
\textit{BK-390 = EM* GGA 50 ($P=4.17429$ $d$):}\\
Listed with a spectral type of B2Ve in SIMBAD and on BeSS, a literature search for this object does not indicate that it is a known EB. No spectra are available on BeSS. This object is listed in a catalog of H$\alpha$ sources \citep{Kohoutek1999}. Given the morphology of the phased light curve and its similarities to a contact binary, this is likely not a classical Be star.  \vspace{2mm} \\ 
\textit{BK-600 = RY Gem = HD 58713 ($P=9.30053$ $d$):}\\
A known EB of Algol type. BeSS spectra show that this is indeed an emission line star. \citet{Plavec1987} classify this as a moderately interacting Algol-type EB consisting of an A0V primary and a K0IV secondary, with clear evidence for circumstellar absorption and emission from circumstellar material around the primary component. The primary eclipse reaches a depth of around 1.5 mag in the KELT bandpass. \vspace{2mm} \\ 
\textit{BK-009 = RW Per = HD 276247 ($P=13.19694$ $d$):}\\
A known EB of Algol type, where the two components are a B9.6e IV-V and a K2 III-IV star \citep{Wilson1988}. Two BeSS spectra show emission with clear double-peaks. \vspace{2mm} \\ 
\textit{BK-480 = V358 Per = HD 13890 ($P=30.28880$ $d$):}\\
BeSS spectra show strong Hydrogen emission with two discernible peaks, and this is classified as a Be star in numerous studies \citep[e.g.][]{Jaschek1982,Steele1999,Clark2000}. A literature search does not indicate that this is a known EB. \vspace{2mm} \\ 
\textit{BK-213 = HD 84511 ($P=33.03207$ $d$):}\\
This star is included in the Jaschek catalog of Be stars, and the single BeSS spectra shows H$\alpha$ in emission. This object shows strong out of eclipse variability in the KELT light curve, compatible with an ellipsoidal variable. At an orbital period of 33.03207 days, at least one component of this system is likely an evolved star with a tenuous atmosphere. A literature search does not indicate that this is a known EB. \vspace{2mm} \\ 
\textit{BK-175 = FY Vel = HD 72754 ($P=33.75389$ $d$):}\\
Many BeSS spectra show an extremely strong, narrow, single-peaked H$\alpha$ emission feature, but this is listed as a shell star in BeSS (B2IIpshe). Simbad lists this as having a spectral type of ``B2(Ia)pe($\_$sh)''. This system is a known $\beta$ Lyrae type EB \citep{Thackeray1970}. \vspace{2mm} \\ 
\textit{BK-013 = EM* MWC 800 ($P=205.99$ $d$):}\\
No BeSS spectra are available for this object. This is included in a catalog of H$\alpha$ emission stars in the Northen Milky Way, which is where the B`e' designation comes from \citep{Kohoutek1999}. Likely an interesting system, but is not a classical Be star judging by the morphology of the KELT light curve when phased to what we presume is the orbital period.

\vspace{60mm}

\newpage
\clearpage

\bibliographystyle{apj}
\bibliography{Main}

\begin{thebibliography}{}
\expandafter\ifx\csname natexlab\endcsname\relax\def\natexlab#1{#1}\fi

\bibitem[{{Baade} {et~al.}(2016){Baade}, {Rivinius}, {Pigulski}, {Carciofi},
  {Martayan}, {Moffat}, {Wade}, {Weiss}, {Grunhut}, {Handler}, {Kuschnig},
  {Mehner}, {Pablo}, {Popowicz}, {Rucinski}, \& {Whittaker}}]{Baade2016}
{Baade}, D., {Rivinius}, T., {Pigulski}, A., {et~al.} 2016, \aap, 588, A56

\bibitem[{{Balona} {et~al.}(2015){Balona}, {Baran},
  {Daszy{\'n}ska-Daszkiewicz}, \& {De Cat}}]{Balona2015}
{Balona}, L.~A., {Baran}, A.~S., {Daszy{\'n}ska-Daszkiewicz}, J., \& {De Cat},
  P. 2015, \mnras, 451, 1445

\bibitem[{{Carciofi}(2011)}]{Carciofi2011}
{Carciofi}, A.~C. 2011, in IAU Symposium, Vol. 272, Active OB Stars: Structure,
  Evolution, Mass Loss, and Critical Limits, ed. C.~{Neiner}, G.~{Wade},
  G.~{Meynet}, \& G.~{Peters}, 325--336

\bibitem[{{Chojnowski} {et~al.}(2015){Chojnowski}, {Whelan}, {Wisniewski},
  {Majewski}, {Hall}, {Shetrone}, {Beaton}, {Burton}, {Damke}, {Eikenberry},
  {Hasselquist}, {Holtzman}, {M{\'e}sz{\'a}ros}, {Nidever}, {Schneider},
  {Wilson}, {Zasowski}, {Bizyaev}, {Brewington}, {Brinkmann}, {Ebelke},
  {Frinchaboy}, {Kinemuchi}, {Malanushenko}, {Malanushenko}, {Marchante},
  {Oravetz}, {Pan}, \& {Simmons}}]{Chojnowski2015}
{Chojnowski}, S.~D., {Whelan}, D.~G., {Wisniewski}, J.~P., {et~al.} 2015, \aj,
  149, 7

\bibitem[{{Clark} \& {Steele}(2000)}]{Clark2000}
{Clark}, J.~S., \& {Steele}, I.~A. 2000, \aaps, 141, 65

\bibitem[{{Cuypers} {et~al.}(1989){Cuypers}, {Balona}, \&
  {Marang}}]{Cuypers1989}
{Cuypers}, J., {Balona}, L.~A., \& {Marang}, F. 1989, \aaps, 81, 151

\bibitem[{{Emilio} {et~al.}(2010){Emilio}, {Andrade}, {Janot-Pacheco},
  {Baglin}, {Guti{\'e}rrez-Soto}, {Su{\'a}rez}, {de Batz}, {Diago}, {Fabregat},
  {Floquet}, {Fr{\'e}mat}, {Huat}, {Hubert}, {Espinosa Lara}, {Leroy},
  {Martayan}, {Neiner}, {Semaan}, \& {Suso}}]{Emilio2010}
{Emilio}, M., {Andrade}, L., {Janot-Pacheco}, E., {et~al.} 2010, \aap, 522, A43

\bibitem[{{Findeisen} {et~al.}(2015){Findeisen}, {Cody}, \&
  {Hillenbrand}}]{Findeisen2015}
{Findeisen}, K., {Cody}, A.~M., \& {Hillenbrand}, L. 2015, \apj, 798, 89

\bibitem[{{Grundstrom} {et~al.}(2011){Grundstrom}, {McSwain}, {Aragona},
  {Boyajian}, {Marsh}, \& {Roettenbacher}}]{Grundstrom2011}
{Grundstrom}, E.~D., {McSwain}, M.~V., {Aragona}, C., {et~al.} 2011, Bulletin
  de la Societe Royale des Sciences de Liege, 80, 371

\bibitem[{{Guti{\'e}rrez-Soto} {et~al.}(2007){Guti{\'e}rrez-Soto}, {Fabregat},
  {Suso}, {Lanzara}, {Garrido}, {Hubert}, \& {Floquet}}]{Gutierrez-Soto2007}
{Guti{\'e}rrez-Soto}, J., {Fabregat}, J., {Suso}, J., {et~al.} 2007, \aap, 476,
  927

\bibitem[{{Guti{\'e}rrez-Soto} {et~al.}(2008){Guti{\'e}rrez-Soto}, {Neiner},
  {Hubert}, {Floquet}, {Huat}, {Diago}, {Fabregat}, {Leroy}, {De Batz},
  {Andrade}, {Emilio}, {Facanha}, {Fremat}, {Janot-Pacheco}, {Martayan}, \&
  {Suso}}]{Gutierrez-Soto2008}
{Guti{\'e}rrez-Soto}, J., {Neiner}, C., {Hubert}, A.-M., {et~al.} 2008,
  Communications in Asteroseismology, 157, 70

\bibitem[{{Hartman}(2012)}]{Hartman2012}
{Hartman}, J. 2012, {VARTOOLS: Light Curve Analysis Program}, Astrophysics
  Source Code Library, ascl:1208.016

\bibitem[{{Haubois} {et~al.}(2012{\natexlab{a}}){Haubois}, {Carciofi},
  {Rivinius}, {Okazaki}, \& {Bjorkman}}]{Haubois2012a}
{Haubois}, X., {Carciofi}, A.~C., {Rivinius}, T., {Okazaki}, A.~T., \&
  {Bjorkman}, J.~E. 2012{\natexlab{a}}, \apj, 756, 156

\bibitem[{{Haubois} {et~al.}(2012{\natexlab{b}}){Haubois}, {Carciofi},
  {Rivinius}, {Okazaki}, \& {Bjorkman}}]{Haubois2012}
{Haubois}, X., {Carciofi}, A.~C., {Rivinius}, T., {Okazaki}, A.~T., \&
  {Bjorkman}, J.~E. 2012{\natexlab{b}}, in Astronomical Society of the Pacific
  Conference Series, Vol. 464, Circumstellar Dynamics at High Resolution, ed.
  A.~C. {Carciofi} \& T.~{Rivinius}, 133

\bibitem[{{Huat} {et~al.}(2009){Huat}, {Hubert}, {Floquet}, {Neiner}, {Saio},
  {Lovekin}, \& {Baudin}}]{Huat2009}
{Huat}, A.-L., {Hubert}, A.-M., {Floquet}, M., {et~al.} 2009, in American
  Institute of Physics Conference Series, Vol. 1170, American Institute of
  Physics Conference Series, ed. J.~A. {Guzik} \& P.~A. {Bradley}, 382--384

\bibitem[{{Hubert}(2007)}]{Hubert2007}
{Hubert}, A.~M. 2007, in Astronomical Society of the Pacific Conference Series,
  Vol. 361, Active OB-Stars: Laboratories for Stellare and Circumstellar
  Physics, ed. A.~T. {Okazaki}, S.~P. {Owocki}, \& S.~{Stefl}, 27

\bibitem[{{Hubert} \& {Floquet}(1998)}]{Hubert1998}
{Hubert}, A.~M., \& {Floquet}, M. 1998, \aap, 335, 565

\bibitem[{{Hutchings} \& {Redman}(1973)}]{Hutchings1973}
{Hutchings}, J.~B., \& {Redman}, R.~O. 1973, \mnras, 163, 219

\bibitem[{{Jaschek} \& {Egret}(1982)}]{Jaschek1982}
{Jaschek}, M., \& {Egret}, D. 1982, in IAU Symposium, Vol.~98, Be Stars, ed.
  M.~{Jaschek} \& H.-G. {Groth}, 261

\bibitem[{{Jones} {et~al.}(2013){Jones}, {Wiegert}, {Tycner}, {Henry}, {Cyr},
  {Halonen}, \& {Muterspaugh}}]{Jones2013}
{Jones}, C.~E., {Wiegert}, P.~A., {Tycner}, C., {et~al.} 2013, \aj, 145, 142

\bibitem[{{Kee} {et~al.}(2014){Kee}, {Owocki}, {Townsend}, \&
  {M{\"u}ller}}]{Kee2014}
{Kee}, N., {Owocki}, S., {Townsend}, R., \& {M{\"u}ller}, H.-R. 2014, ArXiv
  e-prints, arXiv:1412.8511

\bibitem[{{Kee} {et~al.}(2016){Kee}, {Owocki}, \& {Sundqvist}}]{Kee2016}
{Kee}, N.~D., {Owocki}, S., \& {Sundqvist}, J.~O. 2016, \mnras, 458, 2323

\bibitem[{{Kohoutek} \& {Wehmeyer}(1999)}]{Kohoutek1999}
{Kohoutek}, L., \& {Wehmeyer}, R. 1999, \aaps, 134, 255

\bibitem[{{Kov{\'a}cs} {et~al.}(2005){Kov{\'a}cs}, {Bakos}, \&
  {Noyes}}]{Kovacs2005}
{Kov{\'a}cs}, G., {Bakos}, G., \& {Noyes}, R.~W. 2005, \mnras, 356, 557

\bibitem[{{Kroll} \& {Hanuschik}(1997)}]{Kroll1997}
{Kroll}, P., \& {Hanuschik}, R.~W. 1997, in Astronomical Society of the Pacific
  Conference Series, Vol. 121, IAU Colloq. 163: Accretion Phenomena and Related
  Outflows, ed. D.~T. {Wickramasinghe}, G.~V. {Bicknell}, \& L.~{Ferrario}, 494

\bibitem[{{Lee} {et~al.}(1991){Lee}, {Osaki}, \& {Saio}}]{Lee1991}
{Lee}, U., {Osaki}, Y., \& {Saio}, H. 1991, \mnras, 250, 432

\bibitem[{{Lef{\`e}vre} {et~al.}(2009){Lef{\`e}vre}, {Marchenko}, {Moffat}, \&
  {Acker}}]{Lefevre2009}
{Lef{\`e}vre}, L., {Marchenko}, S.~V., {Moffat}, A.~F.~J., \& {Acker}, A. 2009,
  \aap, 507, 1141

\bibitem[{{Lomb}(1976)}]{Lomb1976}
{Lomb}, N.~R. 1976, \apss, 39, 447

\bibitem[{{Luyten}(1938)}]{Luyten1938}
{Luyten}, W.~J. 1938, \mnras, 98, 459

\bibitem[{{Martayan} {et~al.}(2006){Martayan}, {Hubert}, {Floquet}, {Fabregat},
  {Fr{\'e}mat}, {Neiner}, {Stee}, \& {Zorec}}]{Martayan2006}
{Martayan}, C., {Hubert}, A.~M., {Floquet}, M., {et~al.} 2006, \aap, 445, 931

\bibitem[{{Martin} {et~al.}(2005){Martin}, {Fanson}, {Schiminovich},
  {Morrissey}, {Friedman}, {Barlow}, {Conrow}, {Grange}, {Jelinsky},
  {Milliard}, {Siegmund}, {Bianchi}, {Byun}, {Donas}, {Forster}, {Heckman},
  {Lee}, {Madore}, {Malina}, {Neff}, {Rich}, {Small}, {Surber}, {Szalay},
  {Welsh}, \& {Wyder}}]{Martin2005}
{Martin}, D.~C., {Fanson}, J., {Schiminovich}, D., {et~al.} 2005, \apjl, 619,
  L1

\bibitem[{{Martin} {et~al.}(2011){Martin}, {Pringle}, {Tout}, \&
  {Lubow}}]{Martin2011}
{Martin}, R.~G., {Pringle}, J.~E., {Tout}, C.~A., \& {Lubow}, S.~H. 2011,
  \mnras, 416, 2827

\bibitem[{{McSwain} {et~al.}(2009){McSwain}, {Huang}, \& {Gies}}]{McSwain2009}
{McSwain}, M.~V., {Huang}, W., \& {Gies}, D.~R. 2009, \apj, 700, 1216

\bibitem[{{Mennickent} {et~al.}(1998){Mennickent}, {Sterken}, \&
  {Vogt}}]{Mennickent1998}
{Mennickent}, R.~E., {Sterken}, C., \& {Vogt}, N. 1998, \aap, 330, 631

\bibitem[{{Mennickent} {et~al.}(1994){Mennickent}, {Vogt}, \&
  {Sterken}}]{Mennickent1994}
{Mennickent}, R.~E., {Vogt}, N., \& {Sterken}, C. 1994, \aaps, 108

\bibitem[{{Merrill}(1952)}]{Merrill1952}
{Merrill}, P.~W. 1952, \apj, 116, 516

\bibitem[{{Miroshnichenko} {et~al.}(2001){Miroshnichenko}, {Fabregat},
  {Bjorkman}, {Knauth}, {Morrison}, {Tarasov}, {Reig}, {Negueruela}, \&
  {Blay}}]{Miroshnichenko2001}
{Miroshnichenko}, A.~S., {Fabregat}, J., {Bjorkman}, K.~S., {et~al.} 2001,
  \aap, 377, 485

\bibitem[{{Miroshnichenko} {et~al.}(2013){Miroshnichenko}, {Pasechnik},
  {Manset}, {Carciofi}, {Rivinius}, {{\v S}tefl}, {Gvaramadze}, {Ribeiro},
  {Fernando}, {Garrel}, {Knapen}, {Buil}, {Heathcote}, {Pollmann}, {Mauclaire},
  {Thizy}, {Martin}, {Zharikov}, {Okazaki}, {Gandet}, {Eversberg}, \&
  {Reinecke}}]{Miroshnichenko2013}
{Miroshnichenko}, A.~S., {Pasechnik}, A.~V., {Manset}, N., {et~al.} 2013, \apj,
  766, 119

\bibitem[{{Moe} \& {Di Stefano}(2016)}]{Moe2016}
{Moe}, M., \& {Di Stefano}, R. 2016, ArXiv e-prints, arXiv:1606.05347

\bibitem[{{Negueruela} {et~al.}(2004){Negueruela}, {Steele}, \&
  {Bernabeu}}]{Negueruela2004}
{Negueruela}, I., {Steele}, I.~A., \& {Bernabeu}, G. 2004, Astronomische
  Nachrichten, 325, 749

\bibitem[{{Neiner} {et~al.}(2011){Neiner}, {de Batz}, {Cochard}, {Floquet},
  {Mekkas}, \& {Desnoux}}]{Neiner2011}
{Neiner}, C., {de Batz}, B., {Cochard}, F., {et~al.} 2011, \aj, 142, 149

\bibitem[{{Neiner} {et~al.}(2005){Neiner}, {Hubert}, \& {Catala}}]{Neiner2005}
{Neiner}, C., {Hubert}, A.-M., \& {Catala}, C. 2005, \apjs, 156, 237

\bibitem[{{Okazaki}(1991)}]{Okazaki1991}
{Okazaki}, A.~T. 1991, \pasj, 43, 75

\bibitem[{{Oudmaijer} \& {Parr}(2010)}]{Oudmaijer2010}
{Oudmaijer}, R.~D., \& {Parr}, A.~M. 2010, \mnras, 405, 2439

\bibitem[{{{\"O}zdemir} {et~al.}(2003){{\"O}zdemir}, {Mayer}, {Drechsel},
  {Demircan}, \& {Ak}}]{Ozdemir2003}
{{\"O}zdemir}, S., {Mayer}, P., {Drechsel}, H., {Demircan}, O., \& {Ak}, H.
  2003, \aap, 403, 675

\bibitem[{{{\"O}zt{\"u}rk} {et~al.}(2014){{\"O}zt{\"u}rk}, {Soydugan}, \& {{\c
  C}i{\c c}ek}}]{Ozturk2014}
{{\"O}zt{\"u}rk}, O., {Soydugan}, F., \& {{\c C}i{\c c}ek}, C. 2014, \na, 30,
  100

\bibitem[{{Panoglou} {et~al.}(2016){Panoglou}, {Carciofi}, {Vieira}, {Cyr},
  {Jones}, {Okazaki}, \& {Rivinius}}]{Panoglou2016}
{Panoglou}, D., {Carciofi}, A.~C., {Vieira}, R.~G., {et~al.} 2016, \mnras, 461,
  2616

\bibitem[{{Papaloizou} {et~al.}(1992){Papaloizou}, {Savonije}, \&
  {Henrichs}}]{Papaloizou1992}
{Papaloizou}, J.~C., {Savonije}, G.~J., \& {Henrichs}, H.~F. 1992, \aap, 265,
  L45

\bibitem[{{Pepper} {et~al.}(2012){Pepper}, {Kuhn}, {Siverd}, {James}, \&
  {Stassun}}]{Pepper2012}
{Pepper}, J., {Kuhn}, R.~B., {Siverd}, R., {James}, D., \& {Stassun}, K. 2012,
  \pasp, 124, 230

\bibitem[{{Pepper} {et~al.}(2007){Pepper}, {Pogge}, {DePoy}, {Marshall},
  {Stanek}, {Stutz}, {Poindexter}, {Siverd}, {O'Brien}, {Trueblood}, \&
  {Trueblood}}]{Pepper2007}
{Pepper}, J., {Pogge}, R.~W., {DePoy}, D.~L., {et~al.} 2007, \pasp, 119, 923

\bibitem[{{Plavec} \& {Dobias}(1987)}]{Plavec1987}
{Plavec}, M.~J., \& {Dobias}, J.~J. 1987, \aj, 93, 440

\bibitem[{{Rivinius} {et~al.}(1998){Rivinius}, {Baade}, {Stefl}, {Stahl},
  {Wolf}, \& {Kaufer}}]{Rivinius1998}
{Rivinius}, T., {Baade}, D., {Stefl}, S., {et~al.} 1998, \aap, 333, 125

\bibitem[{{Rivinius} {et~al.}(2001){Rivinius}, {Baade}, {{\v S}tefl},
  {Townsend}, {Stahl}, {Wolf}, \& {Kaufer}}]{Rivinius2001}
{Rivinius}, T., {Baade}, D., {{\v S}tefl}, S., {et~al.} 2001, \aap, 369, 1058

\bibitem[{{Rivinius} {et~al.}(2013){Rivinius}, {Carciofi}, \&
  {Martayan}}]{Rivinius2013}
{Rivinius}, T., {Carciofi}, A.~C., \& {Martayan}, C. 2013, \aapr, 21, 69

\bibitem[{{Saio} {et~al.}(2007){Saio}, {Cameron}, {Kuschnig}, {Walker},
  {Matthews}, {Rowe}, {Lee}, {Huber}, {Weiss}, {Guenther}, {Moffat},
  {Rucinski}, \& {Sasselov}}]{Saio2007}
{Saio}, H., {Cameron}, C., {Kuschnig}, R., {et~al.} 2007, \apj, 654, 544

\bibitem[{{Scargle}(1982)}]{Scargle1982}
{Scargle}, J.~D. 1982, \apj, 263, 835

\bibitem[{{Sigut} \& {Patel}(2013)}]{Sigut2013}
{Sigut}, T.~A.~A., \& {Patel}, P. 2013, \apj, 765, 41

\bibitem[{{Skrutskie} {et~al.}(2006){Skrutskie}, {Cutri}, {Stiening},
  {Weinberg}, {Schneider}, {Carpenter}, {Beichman}, {Capps}, {Chester},
  {Elias}, {Huchra}, {Liebert}, {Lonsdale}, {Monet}, {Price}, {Seitzer},
  {Jarrett}, {Kirkpatrick}, {Gizis}, {Howard}, {Evans}, {Fowler}, {Fullmer},
  {Hurt}, {Light}, {Kopan}, {Marsh}, {McCallon}, {Tam}, {Van Dyk}, \&
  {Wheelock}}]{Skrutskie2006}
{Skrutskie}, M.~F., {Cutri}, R.~M., {Stiening}, R., {et~al.} 2006, \aj, 131,
  1163

\bibitem[{{Steele} {et~al.}(1999){Steele}, {Negueruela}, \&
  {Clark}}]{Steele1999}
{Steele}, I.~A., {Negueruela}, I., \& {Clark}, J.~S. 1999, \aaps, 137, 147

\bibitem[{{Sterken} {et~al.}(1994){Sterken}, {Vogt}, \&
  {Mennickent}}]{Sterken1994}
{Sterken}, C., {Vogt}, N., \& {Mennickent}, R. 1994, \aap, 291, 473

\bibitem[{{Sterken} {et~al.}(1996){Sterken}, {Vogt}, \&
  {Mennickent}}]{Sterken1996}
{Sterken}, C., {Vogt}, N., \& {Mennickent}, R.~E. 1996, \aap, 311, 579

\bibitem[{{Thackeray} {et~al.}(1970){Thackeray}, {Alexander}, \&
  {Hill}}]{Thackeray1970}
{Thackeray}, A.~D., {Alexander}, J.~B., \& {Hill}, P.~W. 1970, Information
  Bulletin on Variable Stars, 483

\bibitem[{{Townsend}(2003)}]{Townsend2003}
{Townsend}, R.~H.~D. 2003, \mnras, 340, 1020

\bibitem[{{van Leeuwen}(1997)}]{VanLeeuwen1997}
{van Leeuwen}, F. 1997, \ssr, 81, 201

\bibitem[{{Vesper} \& {Honeycutt}(1993)}]{Vesper1993}
{Vesper}, D.~N., \& {Honeycutt}, R.~K. 1993, \pasp, 105, 731

\bibitem[{{Walker} {et~al.}(2005{\natexlab{a}}){Walker}, {Kuschnig},
  {Matthews}, {Cameron}, {Saio}, {Lee}, {Kambe}, {Masuda}, {Guenther},
  {Moffat}, {Rucinski}, {Sasselov}, \& {Weiss}}]{Walker2005}
{Walker}, G.~A.~H., {Kuschnig}, R., {Matthews}, J.~M., {et~al.}
  2005{\natexlab{a}}, \apjl, 635, L77

\bibitem[{{Walker} {et~al.}(2005{\natexlab{b}}){Walker}, {Kuschnig},
  {Matthews}, {Reegen}, {Kallinger}, {Kambe}, {Saio}, {Harmanec}, {Guenther},
  {Moffat}, {Rucinski}, {Sasselov}, {Weiss}, {Bohlender}, {Bo{\v z}i{\'c}},
  {Hashimoto}, {Koubsk{\'y}}, {Mann}, {Ru{\v z}djak}, {{\v S}koda}, {{\v
  S}lechta}, {Sudar}, {Wolf}, \& {Yang}}]{Walker2005a}
---. 2005{\natexlab{b}}, \apjl, 623, L145

\bibitem[{{Wilson} \& {Plavec}(1988)}]{Wilson1988}
{Wilson}, R.~E., \& {Plavec}, M.~J. 1988, \aj, 95, 1828

\bibitem[{{Wolf} {et~al.}(2006){Wolf}, {Ku{\v c}{\'a}kov{\'a}}, {Kolasa}, {{\v
  S}tastn{\'y}}, {Bozkurt}, {Harmanec}, {Zejda}, {Br{\'a}t}, \&
  {Hornoch}}]{Wolf2006}
{Wolf}, M., {Ku{\v c}{\'a}kov{\'a}}, H., {Kolasa}, M., {et~al.} 2006, \aap,
  456, 1077

\bibitem[{{Wright} {et~al.}(2010){Wright}, {Eisenhardt}, {Mainzer}, {Ressler},
  {Cutri}, {Jarrett}, {Kirkpatrick}, {Padgett}, {McMillan}, {Skrutskie},
  {Stanford}, {Cohen}, {Walker}, {Mather}, {Leisawitz}, {Gautier}, {McLean},
  {Benford}, {Lonsdale}, {Blain}, {Mendez}, {Irace}, {Duval}, {Liu}, {Royer},
  {Heinrichsen}, {Howard}, {Shannon}, {Kendall}, {Walsh}, {Larsen}, {Cardon},
  {Schick}, {Schwalm}, {Abid}, {Fabinsky}, {Naes}, \& {Tsai}}]{Wright2010}
{Wright}, E.~L., {Eisenhardt}, P.~R.~M., {Mainzer}, A.~K., {et~al.} 2010, \aj,
  140, 1868

\end{thebibliography}

\end{document}